\documentclass[english]{article}
\usepackage[latin9]{inputenc}
\usepackage{graphicx}
\usepackage{hyperref}
\usepackage{multirow}
\usepackage[top=2.5cm]{geometry}
\usepackage{authblk}
\usepackage{cite}
\usepackage{comment}
\usepackage[dvipsnames]{xcolor}
\usepackage{amssymb}

\makeatletter

\providecommand{\tabularnewline}{\\}

\makeatother

\usepackage{babel}
\usepackage{amsmath}
\usepackage{slashed}

\title{Exotic Portals for Hadronic Processes: The Gluon-Gluon and Gluon-Quark Portals}
\author[1]{Katherine Schwind\thanks{schwind.44@osu.edu}}
\author[1]{Linda M. Carpenter\thanks{lmc@physics.osu.edu}}

\affil[1]{\emph{Department of Physics, The Ohio State University,
191 W. Woodruff Avenue, Columbus, OH 43210, U.S.A.}}

\begin{document}

\maketitle

\begin{abstract}
In this work we explore all possible light exotic (LEX) particles which are accessible through the gluon-gluon ($g$-$g$) and gluon-quark ($g$-$q$) phenomenological portals up to effective dimension six. We present effective models for these new states, which are directly accessible through gluon and quark fusion processes, and which are guaranteed decays involving Standard Model (SM) quarks and gluons. We note many accessible states in particularly unusual SU(3)$_\text{c}$ representations, which contain large Clebsch-Gordan coefficients. We also  include some interesting dimension seven operators with highly unusual features. We also consider all potential spin values for our LEX states, looking at Lorentz representations and color selection rules. In this, we find that states in various unusual Lorentz representations are accessible. We also find that certain operator types have restrictions on allowed color representations based on spin.

\end{abstract}

\section{Introduction} \label{s1}
In this work, we intend to explore some of the more unusual regions in theory-space. This work is a continuation of our general effort to map the  Beyond the Standard Model (BSM) states which can be accessed through specific phenomenological portals~\cite{Carpenter:2025aaq, Carpenter:2025cys, Carpenter:2024hvp, Carpenter:2024jje}. To this end, we use the techniques of Light Exotics Effective Field Theory (LEX-EFT) to construct effective operators through which exotic particles couple to known Standard Model (SM) states~\cite{Carpenter_2023, Carpenter:2021rkl}. In specifying a particular phenomenological portal, we fix a set of SM states in the desired interaction. We then may then determine all possible BSM states that may be accessed through any effective operator up to the appropriate effective dimension in effective field theory. This technique allows us to determine what kinds of exotic states can be accessed through specific collider production processes. It also gives us a map of possible exotic particle decays. In previous explorations of phenomenological portals, we have found that the menagerie of accessible particles---those reachable through various phenomenological portals---is much greater than those explored in usual BSM model building paradigms. In particular, previously explored phenomenological portals have included exotics with unique combinations of SU(3)$_\text{c}$ and SU(2)$_\text{L}$ quantum numbers. Such considerations are part of a growing body of phenomenological work aimed at opening up collider studies to a broader set of exotic particles. This expansion includes color octets, color sextets~\cite{Carpenter:2023aec, Carpenter:2022qsw, Han:2010rf}, lepto-gluons~\cite{Goncalves-Netto:2013nla, Jelinski:2015epa}, higher charged lepto-quarks~\cite{Blum:2016szr} and others~\cite{Han:2023djl, Han:2025wdy}.

\begin{figure}[h!]
    \centering
    \includegraphics[width=0.5\linewidth]{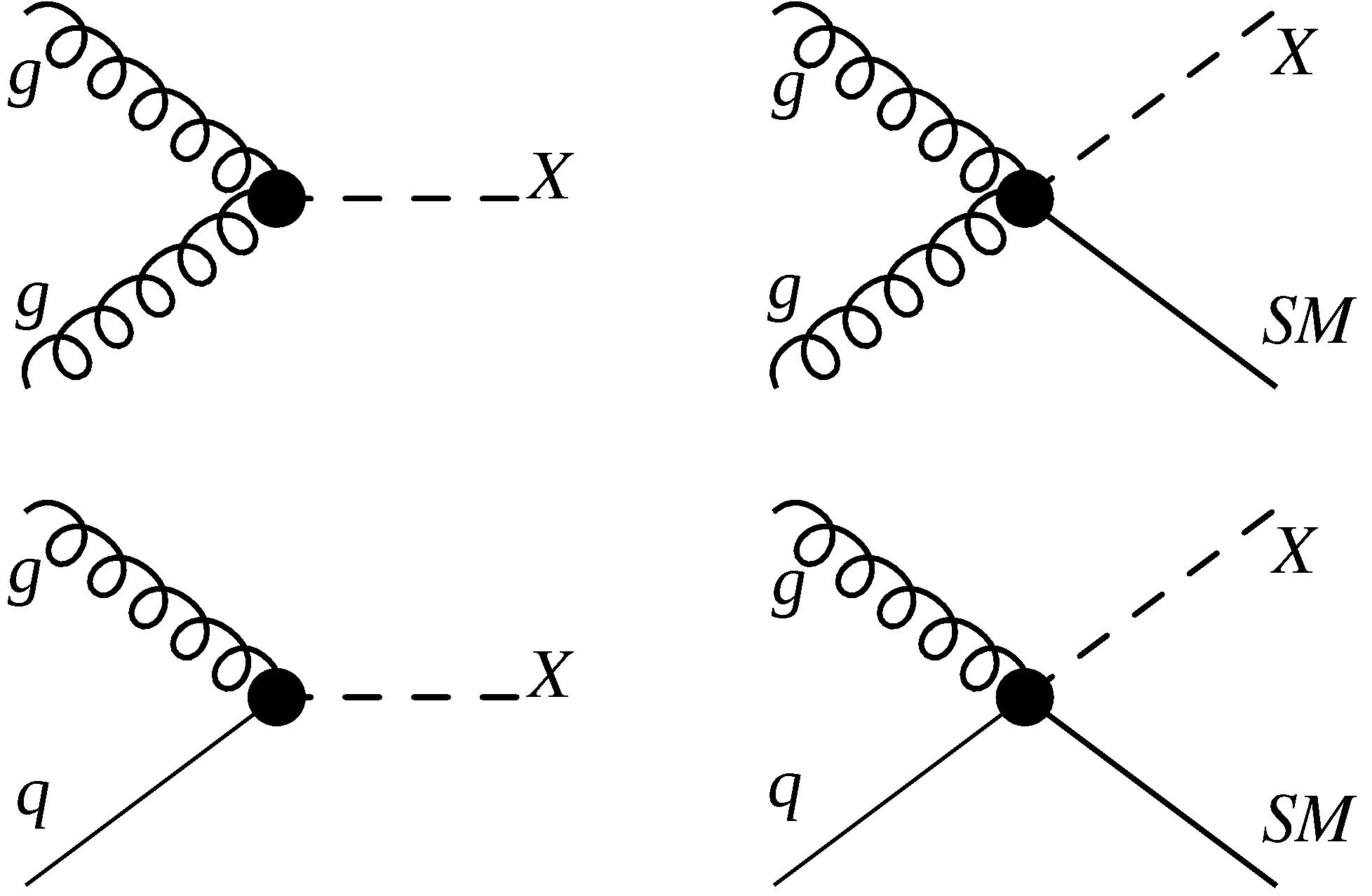}
    \caption{Interaction vertices providing access to BSM states within the gluon-gluon (top row) and gluon-quark (bottom row) portals. $X$ denotes the LEX state, and $SM$ represents any other Standard Model states within an interaction.}
    \label{fig:1}
\end{figure}

In this work, we will focus on two of the most obvious portals for phenomenological work at hadron colliders, the gluon-gluon ($g$-$g$) and gluon-quark ($g$-$q$) portals.  We will determine all possible exotic particles which can be singly produced through interactions containing at at least two SM gluons or at least one quark and one gluon.  A schematic of the effective interactions of interest appears in Fig.~\ref{fig:1}.

We have chosen to focus on interactions that contain a single LEX state for several reasons. First, we wish to explore all models that contain an exotic state that can decay into SM particles at the requisite effective dimension. Second, through various studies of phenomenological portals, we have generically found that, in many EFTs, single production of BSM states at the Large Hadron Collider (LHC)  overtakes pair production from gauge interactions for particle masses above/around 1.5 TeV (for reasonable effective cutoffs). These operators may thus give us the largest potential mass range window for the detection of LEX states in LHC production processes. 

The gluon-gluon and gluon-quark portals allow us to enumerate the exotic particles accessible through some of the most probable LHC processes, gluon fusion and quark-gluon fusion. In this work, we will map out a maximal list of exotics that can be singly produced through these processes. These particular portals also allow us to access the exotic states of highest SU(3)$_\text{c}$ charge at the desired effective dimension in effective field theory. Though exotics of large color charge are certainly phenomenologically accessible, they have not received very much phenomenological attention. In this work, we will explore these higher-charged particles, describe the computation of the SU(3) tensor products for effective operators, and explore some unusual Clebsch-Gordan coefficients.  We will also classify accessible LEX states according to spin. While our previous work on LEX-EFT portals has focused almost entirely on spin 0 and spin 1/2 states, we have made an effort in this work to consider all possible Lorentz representations and spin values for the potential LEX states. While this may seem fanciful, we note that higher spin mesons and baryons exist (for example there is evidence for the $\Delta (2950)$ of spin $15/2$~\cite{ParticleDataGroup:2026mpi}. Where applicable, color-spin selection rules will also be considered. In categorizing effective operators, we will use the SMEFT effective operator convention~\cite{Grzadkowski:2010es} and will specify which operators contain gauge field strength tensors and derivatives. 

We note that, in effective field theories, we always have a choice of basis. Redundant operators may appear through integration by parts, field theory identities, or insertion of equations of motion~\cite{Grzadkowski:2010es, Lehman:2014jma}. In this work, we have chosen the basis in such a way as to maximize the number of operators in the portal. While each specific type of exotic particle in this phenomenological portal may be found in several operators, we have taken care to remove the redundant operators which appear in this portal. Our ultimate choice of basis is not equivalent to the Warsaw basis of SMEFT, but it is instead one that maximizes the number of operators in the portal (and presumably removes other would-be operators in the full EFT). We have chosen this portal-based basis in order to maximize the exploratory power of the phenomenological portal. 

This work proceeds as follows. In Section~\ref{s2}, we discuss some of the choices made throughout this paper in relation to notation of color, Lorentz group, and electroweak charges. In Section~\ref{s3} we enumerate operators involving bosonic LEX states, with an emphasis on scalar states. Section~\ref{s4} examines operators with fermionic (especially spin 1/2) LEX states. Section~\ref{s5} considers some potential LHC and $ep$ collider processes. Section~\ref{s6} concludes.

\section{Conventions and Cast of Characters} \label{s2}
As the portals included in this work both involve two SM states with non-singlet color charge, we must discuss the SU(3)$_\text{c}$ algebra choices that we make throughout this work. The main choice that must be addressed is the use of Clebsch-Gordan coefficients and SU(3)$_\text{c}$ indices. There are two main choices regarding SU(3)$_\text{c}$ indices. We can choose either to write in terms of fundamental indices, or to use only one index, corresponding to the dimension of various states.

In the tables within this work, we will largely follow the notation of Coleman, where an SU(3)$_\text{c}$ object may have $n$ upper and $m$ lower fundamental indices, and is completely symmetric within the upper/lower indices~\cite{Coleman:1965afp}. This state is thus written as
\begin{equation}
\phi^{i_1..i_n}_{j_1...j_m} .
\end{equation}
 The dimension of the SU(3)$_\text{c}$ representation is then calculated using
 \begin{equation}
 dim(n,m)=\frac{1}{2}(n+1)(m+1)(n+m+2).
 \end{equation}
Thus, for example, a state written as $\phi^{i_1 i_2}_{j_1 j_2}$ can be specified as a $(2,2)$ or a  27 of SU(3)$_\text{c}$. Throughout this work, fundamental SU(3)$_\text{c}$ indices will be written using uppercase Roman letters. In general, Clebsch-Gordan coefficients will be listed in terms of fundamental indices. In the sections where adjoint SU(3)$_\text{c}$ indices are used instead, lowercase Roman letters taken from the beginning of the alphabet are used. We note that, if desired, one can convert between SU(3)$_\text{c}$ adjoint and fundamental indices using the SU(3)$_\text{c}$ generators:
\begin{equation}
    X^A_B = X^a (T_a)^A_B.
\end{equation}
Meanwhile, SU(2)$_{\text{L}}$ fundamental indices are written in lowercase Roman letters taken from the middle of the alphabet ($i, j, k...)$. 

In this work, Lorentz indices are indicated with Greek letters. Several sections consider the allowed Lorentz structures available to operators of a certain type. In the interest of brevity, we list here the various Lorentz types which follow throughout this paper, as well as any notable symmetry considerations. We start by listing Lorentz types which can be found within the SM.
\begin{itemize}
    \item $(0,0)$: States with this Lorentz structure are spin 0 scalars. They only have 1 degree of freedom. An example of this within the SM is the Higgs boson.
    \item $(1,0) \oplus (0,1)$: States with this Lorentz structure are spin 1. They are self-dual objects with two anti-symmetric Lorentz indices: $X_{[\mu\nu]}$. These states contain $3+3=6$ degrees of freedom. Any of the SM gauge field strength tensors can be found in this representation.
    \item $(1/2,0) \oplus (0,1/2)$: States with this Lorentz structure are spin 1/2 fermions. They have 4 degrees of freedom. SM Dirac fermions are found in this representation.
    \item $(1/2,1/2)$: This is a spin 1 object that can be written as a Lorentz 4 vector: $X_\mu$. It has 4 degrees of freedom. Covariant derivatives are found in this representation.
\end{itemize} 

We continue by enumerating the various less conventional Lorentz structures which can be found throughout this paper.
\begin{itemize}
    \item $(1,1)$: This is a spin 2 object that can be written with two symmetric indices: $X_{(\mu\nu)}$. It is also traceless. As such, this object has 8 degrees of freedom.
    \item $[(2,0)\oplus(0,2)]$: This is a spin 2 object that can be written with four Lorentz indices as $X_{([\mu\nu][\rho\sigma])}~$\cite{Tybor:1986rg}. This object is antisymmetric under interchange of two indices $\mu\leftrightarrow\nu$ and symmetric under interchange of pairs of indices $\mu\nu\leftrightarrow\rho\sigma$. This originally gives us 21 degrees of freedom: each pair of antisymmetric indices has 6 degrees of freedom, and when we consider the symmetry between pair interchange we find $6 \times 7 /2= 21$ degrees of freedom. We must further satisfy the trace constraint
    \begin{equation}
    g^{\mu\alpha}\phi_{\mu\nu\alpha\beta}=0,
    \end{equation}
    which creates ten additional conditions. We also have 
    \begin{equation}
    \epsilon^{\mu\nu\alpha\beta}\phi_{\mu\nu\alpha\beta},
    \end{equation}
    which is one further condition. This leaves us with 10 degrees of freedom.

    \item $[(3/2,1/2) \oplus(1/2,3/2)]$: This is a spin 2 object. It is written with three Lorentz indices, with two of the indices being antisymmetrized: $\phi_{\mu[\nu\rho]}$. We originally count that this object has 24 degrees of freedom (six for antisymmetrized indices times 4 for the third vector index). From this we must impose the conditions
    \begin{align}
    \begin{split}
    \phi^{\mu}_{\nu\mu}=0  \ \text{(four conditions)}\\
    \epsilon^{\mu\nu\rho\sigma}\phi_{\mu\nu\rho}=0 \ \text{(four conditions)} ,
    \end{split}
    \end{align}
    which  subtract additional decrees of freedom. As such, we find that this object has 16 degrees of freedom~\cite{Lanczos:1962zz}. 
    
    \item $(2,1)  \oplus (1,2)$:  This is a spin 3 object. Presumably, this object may be written with four Lorentz indices: $X_{ (\mu\nu) [\alpha \beta]}$. In this case, the indices $\mu\nu$ are symmetric  and $\alpha\beta$ are antisymmetric. This gives us $9\times6=54$ degrees of freedom. We then impose the further constraints 
    \begin{align}
    \begin{split}
        g^{\mu\alpha}\phi_{ \mu\nu\alpha\beta}=0 \quad
        \text{and} \quad
        \epsilon^{\lambda\nu\alpha\beta}\phi_{ \mu\nu\alpha\beta}=0
    \end{split}
    \end{align}
    to reduce the degrees of freedom to 30. 

    \item  $(3,0)\oplus (0,3)$: This is a spin 3 object.  It can be written as a six-index object with three pairs of indices: $X_{[\mu\nu][\rho\sigma][\alpha\beta]}$~\cite{Damour:1987vm}. It is totally antisymmetric within each pair and totally symmetric under the interchange of any two pairs. When counting degrees of freedom, each antisymmetric pair of indices has 6 degrees of freedom. This yields $8\times7=56$ total degrees of freedom. We then impose the trace constraints 
    \begin{align}
    \begin{split}
    g^{\mu\alpha}\phi_{\mu\nu\alpha\beta\rho\sigma}=0, \\
    \phi_{[\mu\nu\rho]\sigma\alpha\beta} = 0
    \end{split}
    \end{align} 
    where in the second we have antisymmetrized over three indices.  We then get 14 degrees of freedom. 

    \item $(3/2,3/2)$:  This is a spin 3 object. It has three totally symmetric indices, and can be written as $X_{\mu\nu\alpha}$~\cite{DelgadoAcosta:2015ypa}. It contains 16 degrees of freedom.

    \item $(1,1/2)\oplus (1/2,1)$: This is the Rarita-Schwinger field, a spin $3/2$ object with  one Lorentz index: $\psi_{\mu}$. One vector and one Dirac index give 16 degrees of freedom. Then, we set the four trace conditions $\gamma^{\mu}\phi_{\mu}=0$ to give only 12 degrees of freedom.

    \item $[(3/2,0)\oplus(0,3/2)]$: This is a spin $3/2$ object. It can be represented as a fermion that has two antisymmetric Lorentz indices: $\psi_{[\mu\nu]}$~\cite{Kirchbach:1997zb, DelgadoAcosta:2015ypa}. We begin with $6\times4=24$ degrees of freedom. Setting the trace condition 
    \begin{equation} \gamma^{\mu}\psi_{[\mu\nu]}=0
    \end{equation}
    removes 16 of these, leaving 8 degrees of freedom.

    \item $(3/2,1) \oplus (1,3/2)$: This is a spin $5/2$ object. It can be written with two  symmetric Lorentz indices: $\psi_{(\mu\nu)}$. We begin with 40 components, but then impose the trace condition  
    \begin{equation}
        \gamma^{\mu}\psi_{(\mu,\nu)}=0 .
    \end{equation} 
    This applies 16 constraints, leaving us with 24 degrees of freedom.

    \item $(2,1/2) \oplus (1/2,2)$: This is a spin $5/2$ object. It can be written with four Lorentz indices as $\psi_{[\mu \nu] [\rho \sigma]}$. Like the $(2,0)\oplus(0,2)$ state, it is antisymmetric under interchange of two indices $\mu\leftrightarrow\nu$ and symmetric under interchange of pairs of indices $\mu\nu\leftrightarrow\rho\sigma$. We have the further conditions
    \begin{align}
    \epsilon^{\mu\nu\rho\sigma}\psi_{[\mu \nu] [\rho \sigma]}=0 , \\ 
    \gamma^{\mu}\psi_{[\mu \nu] [\rho \sigma]}=0. 
    \end{align}
    This results in 20 degrees of freedom.
\end{itemize}

The last convention we mention, and which is used throughout this paper, is our method of portal specification. In the tables that follow, we indicate if the operator falls into the $g$-$g$ portal, $g$-$q$ portal, or both. Operators in the $g$-$g$ portal will be denoted with a $\diamond$, while operators in the $g$-$q$ portal will be denoted by $\star$. Finally, operators which contribute to both portals will be marked with $\Cap$

\section{Bosonic LEX states} \label{s3}
We begin by enumerating all operators in the gluon-gluon and gluon-quark portals which involve a  bosonic LEX field, denoted in the operators below by $\phi$.

\subsection{Operators of the form $GF\phi$} \label{s3.1}

We first consider operators in the portal that contain a gluon field strength tensor. A single SU(3)$_\text{c}$ field strength tensor in an operator introduces interactions which contain one or two gluons:
\begin{equation}
G_{\mu\nu}= \partial_{\mu} A_{\nu}-\partial_{\nu}A_{\mu}+ ig_3f^{abc}A_{\mu} A_{\nu} .
\end{equation}
As such, an operator with a single SU(3)$_\text{c}$ field strength tensor can automatically be included in the $g$-$g$ portal. However, as field strength tensors have Lorentz indices, we must consider which other fields we need to include in our operators in order to make them Lorentz invariant. 
The simplest operators of this type are of the form $GF\phi$. These contain two field strength tensors and can be found at dimension five. There are no restrictions on which second SM field strength tensor we may choose.  Due to the lack of any fermionic states in operators of this type, these operators will appear in the gluon-gluon portal and not be included in the $g$-$q$ portal.

\subsubsection{Lorentz representations of LEX states}
In operators where a LEX state couples to two SM field strength tensors, we can enumerate the allowed LEX Lorentz representations.  The field strength tensors are in the $(1,0)\oplus(0,1)$ Lorentz representation. As such, to determine the Lorentz representations of the LEX states, we need the tensor product
\begin{align}
[(1,0)\oplus(0,1)] \otimes [(1,0)\oplus(0,1)] = [(2,0)\oplus(0,2)]\oplus [(1,0)\oplus(0,1)]\oplus 2\cdot(0,0)\oplus2\cdot(1,1) .
\end{align}
As such, we find LEX states with spins of 0, 1, and 2.  

\subsubsection{Color-spin selection rules for operators of type $GG\phi$}\label{s3.1.2}

In operators with two SU(3)$_\text{c}$ field strength tensors, we use the SU(3)$_\text{c}$ tensor product
\begin{equation}
    8\otimes 8 = 27 \oplus \overline{10} \oplus 10 \oplus 8 \oplus 8 \oplus 1.
\end{equation} 
The tensor product allows the LEX state to be a 27-plet, 10-plet (decouplet), octet, or singlet under SU(3)$_\text{c}$ in order to make a color singlet operator. However, we will see that not every LEX Lorentz representation can take any color representation. Instead, there are color-spin selection rules resulting from symmetries of the operators.

In color space, we may represent the two gluon  product as an object with two gluon (adjoint) indices: $(GG)^{ab}$. As a two index tensor, we may break this into its symmetric and anti-symmetric parts. These can be written, noting which SU(3)$_\text{c}$ representations are contained in each part, as
\begin{align}
\begin{split}
(GG)^{ab}_{S}\supset 1\oplus8_s\oplus 8_a\oplus 27 \\
 (GG)^{ab}_{A} \supset 8_a\oplus10\oplus \overline{10} .
\end{split}
\end{align}

From here, we can check which Lorentz representations of the LEX state are allowed to appear with which SU(3)$_\text{c}$ representations. We desire to write the LEX states as objects with two gluon indices. In this manner, the color singlet is $\delta^{ab}\phi^{c}_c$. The symmetric and antisymmetric octets are $f^{abc}\phi_c$ and $d^{abc}\phi_c$, respectively. The 27-plet is the two index symmetric object with both a symmetric octet and singlet subtracted: $\phi^{(ab)}-d^{abc}\phi_c-\delta^{ab}\phi^{c}_c$. Finally, the sum of the $10$ and $\overline{10}$ is the antisymmetric two index object minus the anti-symmetric octet: $\phi^{[ab]}-f^{abc}\phi_c$.

 For a LEX state in the $(0,0)$ Lorentz representation, we must contract Lorentz indices between field strength tensors. We note that the product 
 \begin{equation}
 G^a_{\mu\nu}G^{b\mu\nu}
 \end{equation}
 is symmetric under interchange of two gluon indices $a\leftrightarrow b$.
 As such, LEX states with the $(0,0)$ Lorentz representation must be \emph{symmetric} under interchange of gluon indices---meaning the operator exists for scalar LEX states in the $1,8_s$ and $27$ color representations. The decouplet LEX scalars, being anti-symmetric in gluon indices, and the antisymmetric octet cannot couple in this way. 

 A similar symmetry argument holds for pseudo-scalar LEX states coupling to the field strength tensor and its dual. We begin with the structure $G^{ a \mu\nu}\epsilon_{\mu\nu\rho\sigma}G^{b\rho\sigma}$ and interchange gluons $a \leftrightarrow b$. From this, we obtain
 \begin{align}
G^{ b \mu\nu}\epsilon_{\mu\nu\rho\sigma}G^{a \rho\sigma}= G^{ b \rho\sigma}\epsilon_{\rho\sigma\mu\nu}G^{a \mu\nu}=G^{ a \mu\nu}\epsilon_{\mu\nu\rho\sigma}G^{b\rho\sigma} .
 \end{align}
We see that this structure is, in fact, mixed symmetry. The decouplets may therefore couple to a gluon field strength and dual if they are pseudoscalars.

Next, we look at LEX states in the $(1,0)\oplus(0,1)$ representation. Here, we expect operators to have the Lorentz structure
\begin{equation}
G^{a \mu\nu}G^b_{\nu\sigma}\phi_{ab \mu}^{\sigma} .
\end{equation}
As this LEX state is totally antisymmetric in Lorentz indices, it selects the Lorentz part of the two index tensor $G^{a \mu\nu}G^b_{\nu\sigma}$ that is anti-symmetric under $\mu \leftrightarrow \sigma$. This product must therefore also be anti-symmetric under exchange of gluon indices $a\leftrightarrow b$. This means the spin 1 LEX states must be found in the antisymmetric color representations ($8_a$, 10, $\overline{10}$) in order to couple to two gluon field strength tensors.

A spin 2 LEX state in the $(1,1)$  Lorentz representation results in operators with the Lorentz structure 
\begin{equation}
     G^{a \mu\nu}G^b_{\nu\sigma} \phi^\sigma_\mu .
\end{equation}
The symmetry of the LEX state then picks out the symmetric part of the product $G^{a \mu\nu}G^b_{\nu\sigma}$ under $\mu \leftrightarrow \sigma$. This symmetric part is invariant under interchange of $a$ and $b$. As such, we see it must couple to LEX states in the symmetric products of SU(3) (the 1, $8_s$, and 27).

Finally, we check the LEX field in the $(2,0)\oplus(0,2)$ Lorentz representation. Checking our operator $G^{a \mu\nu}G^{b \rho\sigma} \phi_{\mu\nu\rho\sigma}$ under gluon index exchange, we find
\begin{align}
\begin{split}
G^{b \mu\nu}G^{a \rho\sigma} \phi_{\mu\nu\rho\sigma} 
= G^{a \mu\nu}G^{b \rho\sigma} \phi_{\rho\sigma\mu\nu}= G^{a \mu\nu}G^{b \rho\sigma} \phi_{\mu\nu\rho\sigma} ,
\end{split}
\end{align} 
where in the first equality we exchanged indices $\mu \leftrightarrow \rho$ and $\nu \leftrightarrow \sigma$ and in the second equality we used the symmetry of  $\phi_{\mu\nu\rho\sigma}$ under interchange of pairs of indices. This shows that $\phi_{\mu\nu\rho\sigma}$ is symmetric under interchange of its two gluon indices, so it must be in the symmetric 1, $8_s$, or 27 SU(3)$_\text{c}$ representations.

We summarize the results of this section in Table 1. We include columns for the operator itself, the allowed SU(3)$_\text{c}$ charge of the LEX state, and the Lorentz representation of the LEX state. Each of these operators are in the gluon-gluon portal, and all exotics in Table 1 are SU(2)$_\text{L}\times$U(1)$_Y$ singlets.
\begin{table}[h!]
\begin{center}
\begin{tabular}{|c|c|c|}
\hline 
Operators & SU(3)$_{\text{c}}$ & Lorentz Representation \tabularnewline
\hline 
\hline
$G^{\mu\nu }_{} G_{\mu \nu }^{} \phi $ & 1, $8_s$, 27 & $(0,0)$ \tabularnewline
\hline 
$G^{\mu\nu }_{} \widetilde{G}_{\mu \nu }^{} \widetilde{\phi} $ & 10, $\overline{10}$ $8_a$ & $(0,0)_{-}$ \tabularnewline
\hline 
$G^{\mu\nu }_{} \widetilde{G}_{\mu \nu }^{} \widetilde{\phi} $ & 1, $8_a$,27 & $(0,0)_{-}$ \tabularnewline
\hline 
$G^{ \mu\nu}G_{\nu\sigma}\phi_{ \mu}^{\sigma}$, & $8_a$, 10, $\overline{10}$ & $(1,0)\oplus(0,1)$ \tabularnewline
\hline 
$G^{\mu\nu}G_{\nu\sigma}\phi_{ \mu}^{\sigma}$ & 1, $8_s$, 27 & $(1,1)$ \tabularnewline
\hline 
$G^{ \mu\nu}G^{ \rho\sigma} \phi_{\mu\nu\rho\sigma}$ & 1, $8_s$, 27 & $(2,0)\oplus(0,2)$ \tabularnewline
\hline 
\end{tabular}
\end{center}
\caption{Operators of type $GG\phi$, where $\phi$ may have a range of Lorentz structures.}
\end{table}

\subsubsection{Enumeration of Operators of the type $GF\phi$} \label{s3.1.3}
Alternatively, we may couple an exotic state to one SU(3)$_\text{c}$ and one other SM field strength tensor. In this case the LEX fields are simple color octets.  In Table~\ref{tab:1}, we display all dimension five operators of type $GF\phi$, where $\phi$ is a spin 0 scalar state. 
\begin{table}[h!]
\begin{center}
\begin{tabular}{|c|c|c|c|}
\hline 
CG Coefficients & Operators & (SU(3)$_{\text{c}}$, SU(2)$_{\text{L}}$, U(1)$_Y$) & Portal \tabularnewline
\hline 
\hline
${C_{8,8,27}}^{BDGH}_{ ACEF}$ & $G^{\mu\nu A}_{B} G_{\mu \nu D}^{C} \phi^{EF}_{GH} $ & ($27,1,0$) & $\diamond$ \tabularnewline
\hline 
${C_{8,8,8}}^{BDF}_{ACE}$ & $G^{\mu\nu A}_{B} G_{\mu \nu D}^{C} \phi^{E}_{F} $ & ($8_s,1,0$) & $\diamond$ \tabularnewline
\hline 
${C_{8,8}}^{BD}_{AC}$ & $G^{\mu\nu A}_{B} G^C_{\mu \nu D} \phi $ & ($1,1,0$) & $\diamond$ \tabularnewline
\hline
${C_{8,8}}^{BD}_{AC}$ & $G^{\mu\nu A}_B W_{\mu \nu}^{ij} \phi^C_{ ij D}$ & ($8,3,0$) & $\diamond$ \tabularnewline
 \hline
${C_{8,8}}^{BD}_{AC}$ & $G^{\mu\nu A}_B B_{\mu \nu} \phi^C_{D}$ & ($8,1,0$) & $\diamond$ \tabularnewline
 \hline
\end{tabular}
\end{center}
\caption{\label{tab:1}Dimension 5 operators of type $GF\phi$, where $\phi$ is a scalar state.}
\end{table}
We list the Clebsch-Gordan coefficients and operators in the first 2 columns. In the third column, we state the SU(3)$_\text{c}\times$SU(2)$_\text{L}\times$U(1)$_Y$ quantum numbers of the accessible LEX scalar. The last column specifies the portal. We note with respect to the Clebsch-Gordan coefficients that there may be multiple index contraction choices available for a given operator. However, we do not delve into this within this work, but instead list a generic object that merely has the correct number of upper/lower SU(3)$_\text{c}$ indices. We also note that the first three operators of Table~\ref{tab:1} have technically been addressed within Table 1. However, we choose to list them again in Table~\ref{tab:1} for the sake of completeness of the enumeration of operators involving scalar LEX states. Within Table~\ref{tab:1}, all of the LEX states have no hypercharge. In the operator containing the SU(2)$_\text{L}$ field strength tensor, the LEX state is an SU(2)$_\text{L}$ triplet. This is the scalar bi-adjoint state, and this operator has been explored in previous work by the authors~\cite{Carpenter:2024qti, Carpenter:2024hvp}. All other LEX states are SU(2)$_\text{L}$ singlets.

\begin{figure}[h!]
    \centering
    \includegraphics[width=0.3\linewidth]{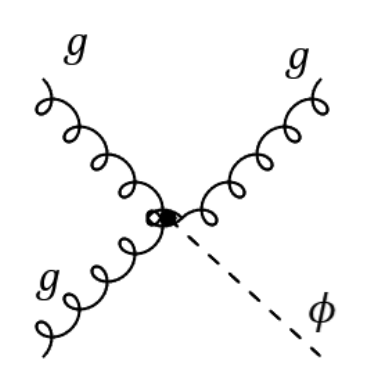}
    \includegraphics[width=0.3\linewidth]{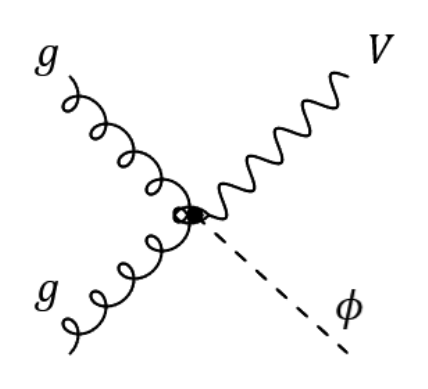}
    \caption{Interaction vertices described in Table~\ref{tab:1}.}
    \label{fig:2}
\end{figure}

The operators of Table 1 contain 2, 3, and 4 gluon vertices that couple to a singlet, octet, decouplet, or 27-plet LEX state. The last six operators in in Table~\ref{tab:1} contain an octet LEX state, one or two gluons, and one electroweak gauge boson.   
In Fig.~\ref{fig:2}, we display two potential interaction vertices from the above tables, both of which involve a total of three gauge bosons.

Operators in Table~\ref{tab:1} carry interesting phenomenological signatures. For example, preliminary explorations of the last operator type appear in \cite{Carpenter:2015gua}. This operator allows the decay of a  color octet to a photon-gluon resonance. Color octet collider processes may contain multiple electroweak gauge bosons through the production and decay processes $gg\rightarrow V+\phi \rightarrow V+(Vg)$.


 
\subsection{Operators of the form $GF\phi H$}
At dimension six, we can add a Higgs scalar to the operators of Table 1 and  Table~\ref{tab:1}. We note that adding the Higgs to the operator does not change the color-spin selection rules in Section~\ref{s3.1.2}. In Table~\ref{tab:2}, we display the dimension six operators of type $GFH\phi$. For brevity we write only operators where $\phi$ is a CP even scalar, with other Lorentz structures following from Table 1. 
\begin{table}[h!]
\begin{center}
\begin{tabular}{|c|c|c|}
\hline 
Operators & (SU(3)$_{\text{c}}$, SU(2)$_{\text{L}}$, U(1)$_Y$) & Portal \tabularnewline
\hline 
\hline 
{$G^{\mu\nu} G_{\mu \nu} \phi^i H_i$} & ($27,2,-1/2$), ($8,2,-1/2$), ($1,2,-1/2$) & {$\diamond$} \tabularnewline
\hline 
$G^{\mu\nu} W_{\mu \nu j}^{i} \phi^{j} H_i$ & ($8,{2},-1/2$) & $\diamond$ \tabularnewline
\hline 
$G^{\mu\nu} W_{\mu \nu j}^{i} \phi_{i}^{jk} H_k$ & ($8,4,-1/2$) & $\diamond$
 \tabularnewline
 \hline
$G^{\mu\nu} B_{\mu \nu } \phi^{i} H_i$ & ($8,{2},-1/2$) & $\diamond$ \tabularnewline
\hline 
\end{tabular}
\end{center}
\caption{\label{tab:2}Dimension 6 operators of type $GFH\phi$, where $\phi$ are CP even scalar LEX states only. SU(3)$_\text{c}$ indices are not written out.}
\end{table}
For these operators, the contraction of color indices works identically to those in Section~\ref{s3.1}; as such, we have chosen not to explicitly write out the color indices in Table~\ref{tab:2}. 

Within Table~\ref{tab:2}, we find LEX states that are electroweak doublets or quadruplets. In all operators without $W_{\mu\nu}$, the SU(2)$_\text{L}$ field strength tensor, the LEX scalar must be an electroweak doublet. In this case, there are two possible ways to write this SU(2)$_\text{L}$ contraction. First, we may write the LEX state as an anti-fundamental of SU(2)$_\text{L}$: $\overline{2}$. In this case, the LEX state has an upper SU(2)$_\text{L}$ index, and the contraction is written as $\phi^iH_i$. Alternatively, we may write the LEX state as a fundamental, in which case the operator is $\phi_i H_j\epsilon^{ij}$. Ultimately, both choices describe the same physics, and we only need to include one of these options in Table~\ref{tab:2}.
In the case where the operator of type $GFH\phi$ contains $W_{\mu\nu}$, we look to the SU(2)$_\text{L}$ product rule $2\otimes3 = 2\oplus 4$. In this case, the LEX state must be an SU(2)$_\text{L}$ doublet or quadruplet. The color octet electroweak doublet found here is the well-known Manohar-Wise field~\cite{Manohar:2006ga}.
The LEX states in all operators of this type must have hypercharge $-1/2$.

\begin{figure}[h!]
    \centering
    \includegraphics[width=0.3\linewidth]{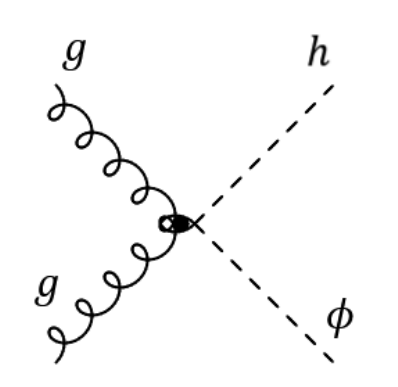}
    \caption{Interaction vertex attainable through the first operator in Table~\ref{tab:2}.}
    \label{fig:3}
\end{figure}

The doublet scalar LEX fields found in Table~\ref{tab:2} contain states of neutral and unit charge: $\phi= (\phi^0,\phi^-)$. For the operators in Table~\ref{tab:2}, if the Higgs vacuum expectation value (VEV) is inserted, this will lead to effective dimension five operators of effective cutoff $v/\lambda^2$ with the same particles at the vertices as in Section~\ref{s3.1.3}.  If instead a physical Higgs field appears, 
then the operators of Table~\ref{tab:2} will produce four, five, or six particle vertices. The four particle vertex that originates from the first operator in Table~\ref{tab:2} involves two gluons, a Higgs, and the neutral LEX state. This interaction vertex is shown in Fig.~\ref{fig:3}.
 Operators involving two SU(3)$_\text{c}$ field strength tensors will contribute to gluon fusion processes with Higgs-LEX associated production through $gg\rightarrow h + \phi^0$  for the various colored LEX particles.  Furthermore, the operator  guarantees the decay of the neutral LEX state to two gluons through the  process $\phi^0\rightarrow g g $, once the Higgs VEV is inserted. If the di-gluon is the only available decay channel for the neutral LEX state, then this operator will lead to a strong di-jet resonance signature through the process $gg\rightarrow \phi^0\rightarrow gg$. However, if the charged and neutral states of the multiplet are split, there is the possibility of more complex signatures through electroweak cascade of the neutral state as discussed below.

To see decays involving the charged state, we invoke the Goldstone-Boson equivalence theorem (EFT implementation has been discussed in SMEFT~\cite{Lewis:2026zug}). The  goldstones within the Higgs multiplet are eaten by the longitudinal modes of the $W$ and $Z$ bosons. We expect LEX couplings in our EFT to $W_L$ and $Z_L$. We thus have a three body decay of the charged state: $\phi^- \rightarrow g g W^-_L$. The masses of the charged and neutral states may be split by a renormalizable dimension four operator, $\lambda|\phi^iH_i|^2$. The charged and neutral LEX masses may  be split to either let  the charged or neutral components be the lightest state. The LEX multiplets  have the additional purely electroweak decay channel $\phi^{\pm/0}\rightarrow W^{\pm/\mp} \phi^{0/\pm} $ through an on- or off-shell $W$ boson. If the charged state is lighter, then a strong di-jet resonance signature for the LEX states may be avoided. For example, we may have single production of the neutral state, which then decays into the charged state.

\begin{figure}[h!] 
    \centering
    \includegraphics[width=0.3\linewidth]{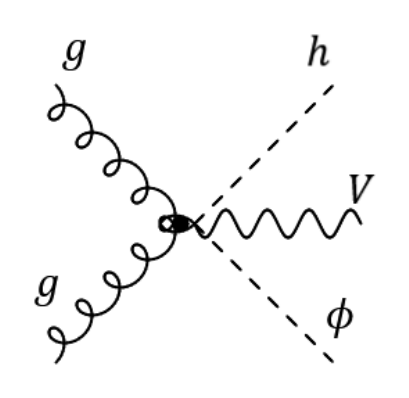} 
    \includegraphics[width=0.4\linewidth]{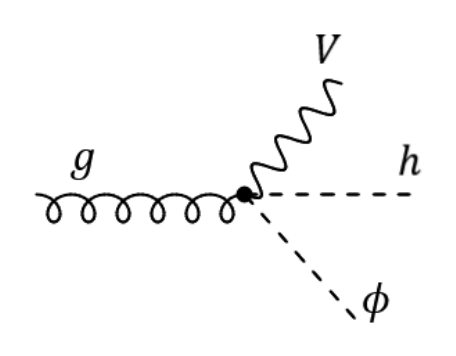} 
    \caption{Interaction vertex attainable through operators 2--4 in Table~\ref{tab:2}.}
    \label{fig:4}
\end{figure}

Operators in Table~\ref{tab:2} will produce vertices with one or two gluons, a Higgs state, one or two electroweak gauge bosons, and a LEX scalar. Some of these potential interactions are shown in Fig.~\ref{fig:4}.
These operators contribute to production processes for the neutral and unit charged LEX $\phi$ states. With insertion of the Higgs VEV, we have the the 2--2 production processes $gg/qq \rightarrow \phi_0 Z/\gamma $ and $gg/ qq\rightarrow W^{\pm/ \mp}\phi^{\mp /\pm} $. This is an interesting electroweak boson associated production channel. 

We can also see that the third operator in Table~\ref{tab:2} contains an SU(2)$_\text{L}$ quadruplet color octet with charge states $(1,0,-1,-2)$. With the Goldstone-Boson equivalence theorem, we can see that the there will be interesting multi electroweak boson production/decay channels for the doubly charged quaadruplet state.

Finally, the operators allow the two body decays of the doublet and quadruplet states to a gluon and electroweak gauge boson. Specifically, the neutral state may decay as $\phi_0\rightarrow g Z/\gamma$. The charged state may similarly decay through the process $\phi^{\mp/\pm}\rightarrow g W^{\mp/\pm}$. 

\subsection{ Operators for Spin 1 LEX states: $G\phi$, $G\phi H$, $G\phi H H$, and $G\phi HHH$ }
For a LEX field in the $(1,0)\oplus(0,1)$ Lorentz representation, we have some additional operators we can include. We may couple this at dimension four to solely a gluon field-strength tensor. We may also include  up to three Higgs fields at dimension six. In the case of one Higgs field, we would find the operator $H G_{\mu\nu}\phi^{\mu\nu}$, where the LEX state has quantum numbers $(8,2,1/2)$. We would also find the operator
$H^{\dagger}G_{\mu\nu}\phi^{\mu\nu}$
with a $(8,2,-1/2)$ LEX state. In the case of two Higgs fields, we also have the operators
$  G_{\mu\nu}H^{\dagger}\phi^{\mu\nu}H$ and $ H^{\dagger} HG_{\mu\nu}\phi^{\mu\nu}$
for $(8,3,0)$ and $(8,1,0)$ LEX states.
Finally, we find the dimension 6 operator forms $ G_{\mu\nu}\phi^{\mu\nu}HHH$ with LEX state $(8,4,-3/2)$ and $ G_{\mu\nu}\phi^{\mu\nu}HHH^{\dagger} $ with LEX state $(8,2,-1/2)$.

\subsection{Operators of form $Gff\phi$} \label{s3.3}
We next choose to examine operators which contain the SU(3)$_\text{c}$ field strength tensor, a LEX scalar, and two SM fermions. There are no restrictions on which SM fermions may be chosen in the gluon-gluon portal. However, only operators where at least one of the SM fermions is a quark will be part of the gluon-quark portal.  All operators of this type are dimension six. These operators are shown in Table~\ref{tab:3} and Table~\ref{tab:4} for LEX states that are Lorentz scalars. We discuss LEX states of higher spin further on, in Section~\ref{s3.4.1}.
\begin{table}[h!]
\begin{center}
\begin{tabular}{|c|c|c|c|}
\hline 
CG Coefficients & Operators & (SU(3)$_{\text{c}}$, SU(2)$_{\text{L}}$, U(1)$_Y$) & Portal \tabularnewline
\hline 
\hline 
${C_{\overline{15},8,3}}^{BEF}_{ ACD}$ & $G^{\mu\nu A}_B \overline{Q_L^c}^C_i \sigma_{\mu\nu} L_j \phi^{ij D}_{EF}$ & ($\overline{15},3,1/3$) & $\Cap$  \tabularnewline 
\hline 
${C_{\overline{15},8,3}}^{BEF}_{ ACD}$ & $G^{\mu\nu A}_B \overline{Q_L^c}^{iC} \sigma_{\mu\nu} L_i \phi^D_{EF}$ & ($\overline{15},1,1/3$) & $\Cap$  \tabularnewline 
\hline 
${C_{{15},8,\overline{3}}}^{BCF}_{ ADE}$ & $G^{\mu\nu A}_B \overline{Q_L}^{i}_C \sigma_{\mu\nu} \ell \phi^{DE}_{i F}$ & (${15},2,7/6$) & $\Cap$  \tabularnewline 
\hline 
${C_{{15},8,\overline{3}}}^{BCF}_{ ADE}$ & $G^{\mu\nu A}_B \overline{u}_C \sigma_{\mu\nu} L_i \phi^{iDE}_F$ & ($15,2,7/6$)  & $\Cap$ \tabularnewline 
\hline 
${C_{\overline{15},8,3}}^{BEF}_{ ACD}$ & $G^{\mu\nu A}_B \overline{u^c}^C \sigma_{\mu\nu} \ell \phi^D_{EF}$ &  ($\overline{15},1,1/3$) & $\Cap$ \tabularnewline 
\hline 
${C_{{15},8,\overline{3}}}^{BCF}_{ ADE}$ & $G^{\mu\nu A}_B \overline{d}_C \sigma_{\mu\nu} L_i \phi^{iDE}_{F}$ & ($15,2,1/6$)  & $\Cap$ \tabularnewline 
\hline 
${C_{\overline{15},8,3}}^{BEF}_{ ACD}$& $G^{\mu\nu A}_B \overline{d^c}^C \sigma_{\mu\nu} \ell \phi^D_{EF}$ &  ($\overline{15},1,4/3$) & $\Cap$ \tabularnewline 
\hline 

${C_{8,6,3}}^{B}_{ ACDE}$ & $G^{\mu\nu A}_B \overline{Q_L^c}^C_i \sigma_{\mu\nu} L_j \phi^{ij DE}$ & ($6,3,1/3$) & $\Cap$  \tabularnewline 
\hline 
${C_{8,6,3}}^{B}_{ ACDE}$ & $G^{\mu\nu A}_B \overline{Q_L^c}^{iC} \sigma_{\mu\nu} L_i \phi^{DE}$ & ($6,1,1/3$) & $\Cap$  \tabularnewline 
\hline 
${C_{8,\overline{6},\overline{3}}}^{BCDE}_{ A}$ & $G^{\mu\nu A}_B \overline{Q_L}^{i}_C \sigma_{\mu\nu} \ell \phi_{iDE}$ & ($\overline{6},2,7/6$) & $\Cap$  \tabularnewline 
\hline 
${C_{8,\overline{6},\overline{3}}}^{BCDE}_{ A}$ & $G^{\mu\nu A}_B \overline{u}_C \sigma_{\mu\nu} L_i \phi^{i}_{DE}$ & ($\overline{6},2,7/6$)  & $\Cap$ \tabularnewline 
\hline 
${C_{8,6,3}}^{B}_{ ACDE}$ & $G^{\mu\nu A}_B \overline{u^c}^C \sigma_{\mu\nu} \ell \phi^{DE}$ &  ($6,1,1/3$) & $\Cap$ \tabularnewline 
\hline 
${C_{8,\overline{6},\overline{3}}}^{BCDE}_{ A}$ & $G^{\mu\nu A}_B \overline{d}_C \sigma_{\mu\nu} L_i \phi^{i}_{DE}$ & ($\overline{6},2,1/6$)  & $\Cap$ \tabularnewline 
\hline 
${C_{8,6,3}}^{B}_{ ACDE}$& $G^{\mu\nu A}_B \overline{d^c}^C \sigma_{\mu\nu} \ell \phi^{DE}$ &  ($6,1,4/3$) & $\Cap$ \tabularnewline 
\hline 

${C_{8,\overline{3},3}}^{BD}_{ AC}$ & $G^{\mu\nu A}_B \overline{Q_L^c}^C_i \sigma_{\mu\nu} L_j \phi^{ij}_D$ & ($\overline{3},3,1/3$) & $\Cap$  \tabularnewline 
\hline 
${C_{8,\overline{3},3}}^{BD}_{ AC}$ & $G^{\mu\nu A}_B \overline{Q_L^c}^{iC} \sigma_{\mu\nu} L_i \phi_D$ & ($\overline{3},1,1/3$) & $\Cap$  \tabularnewline 
\hline 
${C_{8,{3},3}}^{BC}_{ AD}$ & $G^{\mu\nu A}_B \overline{Q_L}_{iC} \sigma_{\mu\nu} \ell \phi_{i}^D$ & (${3},2,7/6$) & $\Cap$  \tabularnewline 
\hline 
${C_{8,\overline{3},3}}^{BC}_{ AD}$ & $G^{\mu\nu A}_B \overline{u}_C \sigma_{\mu\nu} L_i \phi^{iD}$ & ($3,2,7/6$)  & $\Cap$ \tabularnewline 
\hline 
${C_{8,\overline{3},3}}^{BD}_{ AC}$ & $G^{\mu\nu A}_B \overline{u^c}^C \sigma_{\mu\nu} \ell \phi_D$ &  ($\overline{3},1,1/3$) & $\Cap$ \tabularnewline 
\hline 
${C_{8,\overline{3},3}}^{BC}_{ AD}$ & $G^{\mu\nu A}_B \overline{d}_C \sigma_{\mu\nu} L_i \phi^{iD}$ & ($3,2,1/6$)  & $\Cap$ \tabularnewline 
\hline 
${C_{8,\overline{3},3}}^{BD}_{ AC}$& $G^{\mu\nu A}_B \overline{d^c}^C \sigma_{\mu\nu} \ell \phi_D$ &  ($\overline{3},1,4/3$) & $\Cap$ \tabularnewline 
\hline 

${C_{8,8}}^{BD}_{AC}$ & $G^{\mu\nu A}_B \overline{L^c}_i \sigma_{\mu\nu} L_j \phi^{ij C}_D$ & ($8,3,1$)  & $\diamond$ \tabularnewline 
\hline 
${C_{8,8}}^{BD}_{AC}$ & $G^{\mu\nu A}_B \overline{L^c}_i \sigma_{\mu\nu} L_i \phi^C_D$ & ($8,1,1$)  & $\diamond$ \tabularnewline 
\hline 
${C_{8,8}}^{BD}_{AC}$ & $G^{\mu\nu A}_B \overline{L}^i \sigma_{\mu\nu} \ell \phi^C_{iD}$ & ($8,2,1/2$)  & $\diamond$ \tabularnewline 
\hline 
${C_{8,8}}^{BD}_{AC}$ & $G^{\mu\nu A}_B \overline{\ell^c} \sigma_{\mu\nu} \ell \phi^C_D$ &  ($8,1,2$)  & $\diamond$ \tabularnewline 
\hline 
\end{tabular}
\end{center}
\caption{\label{tab:3}Dimension 6 operators of type $Gff\phi$, where at least one of the SM fermions is not a quark and $\phi$ is a scalar state.}
\end{table}

\begin{table}[h!]
\begin{center}
\begin{tabular}{|c|c|c|c|}
\hline 
CG Coefficients & Operators & (SU(3)$_{\text{c}}$, SU(2)$_{\text{L}}$, U(1)$_Y$) & Portal \tabularnewline
\hline 
\hline 
${C_{8,3,3,\overline{24}}}^{BFGH}_{ ACDE}$ & $G^{\mu\nu A}_B \overline{Q_L^c}^C_i \sigma_{\mu\nu} Q_{Lj}^D \phi^{ij E}_{FGH}$ & ($\overline{24},3,-1/3$) & $\Cap$ \tabularnewline 
\hline  
${C_{8,3,3,15}}^{BG}_{ACDEF}$ & $G^{\mu\nu A}_B \overline{Q_L^c}^C_i\sigma_{\mu\nu} Q^D_{Lj} \phi^{ij EF}_G$ & ($15,3,-1/3$) & $\Cap$ \tabularnewline 
\hline  
${C_{8,3,3,\overline{6}}}^{BEF}_{ACD}$  & $G^{\mu\nu A}_B \overline{Q_L^c}^C_i\sigma_{\mu\nu} Q_{Lj}^D \phi^{ij}_{EF}$ & ($\overline{6},3,-1/3$) & $\Cap$ \tabularnewline 
\hline  
${C_{8,3,3,3}}^{B}_{ACDE}$ & $G^{\mu\nu A}_B \overline{Q_L^c}^C_i\sigma_{\mu\nu} Q_{Lj}^D \phi^{ij E}$ & ($3,3,-1/3$) & $\Cap$ \tabularnewline 
\hline
${C_{8,3,3,\overline{24}}}^{BFGH}_{ACDE}$ & $G^{\mu\nu A}_B \overline{Q_L^c}^{iC}\sigma_{\mu\nu} Q_{Li}^D \phi^E_{FGH}$ & ($\overline{24},1,-1/3$) & $\Cap$ \tabularnewline 
\hline  
${C_{8,3,3,15}}^{BG}_{ACDEF}$ & $G^{\mu\nu A}_B \overline{Q_L^c}^{iC}\sigma_{\mu\nu} Q_{Li}^D \phi^{EF}_G$ & ($15,1,-1/3$) & $\Cap$ \tabularnewline 
\hline  
${C_{8,3,3,\overline{6}}}^{BEF}_{ACD}$  & $G^{\mu\nu A}_B \overline{Q_L^c}^{iC}\sigma_{\mu\nu} Q_{Li}^D \phi_{EF}$ & ($\overline{6},1,-1/3$) & $\Cap$ \tabularnewline 
\hline  
${C_{8,3,3,3}}^{B}_{ACDE}$ & $G^{\mu\nu A}_B \overline{Q_L^c}^{iC}\sigma_{\mu\nu} Q_{Li}^D \phi^E$ & ($3,1,-1/3$) & $\Cap$ \tabularnewline 
\hline
${C_{8,\overline{3},3,27}}^{BCGH}_{ ADEF}$ & $G^{\mu\nu A}_B \overline{Q_L}^i_C \sigma_{\mu\nu} u^D \phi^{EF}_{i GH}$ & ($27,2,-1/2$)  & $\Cap$ \tabularnewline 
\hline   
${C_{8,\overline{3},3,10}}^{BC}_{ ADEFG}$ & $G^{\mu\nu A}_B \overline{Q_L}^i_C \sigma_{\mu\nu} u^D \phi_i^{EFG}$ & ($10,2,-1/2$)  & $\Cap$ \tabularnewline 
\hline  
${C_{8,\overline{3},3,\overline{10}}}^{BCEFG}_{ AD}$ & $G^{\mu\nu A}_B \overline{Q_L}^i_C \sigma_{\mu\nu} u^D \phi_{iEFG}$ & ($\overline{10},2,-1/2$)  & $\Cap$ \tabularnewline 
\hline  
${C_{8,\overline{3},3,8}}^{BCF}_{ ADE}$ & $G^{\mu\nu A}_B \overline{Q_L}^i_C \sigma_{\mu\nu} u^D \phi_{iF}^{E}$ & ($8,2,-1/2$)  & $\Cap$ \tabularnewline 
\hline  
${C_{8,\overline{3},3}}^{BC}_{ AD}$ & $G^{\mu\nu A}_B \overline{Q_L}^i_C \sigma_{\mu\nu} u^D \phi_{i}$ & ($1,2,-1/2$)  & $\Cap$ \tabularnewline 
\hline 
${C_{8,\overline{3},3,27}}^{BCGH}_{ ADEF}$ & $G^{\mu\nu A}_B \overline{Q_L}^i_C \sigma_{\mu\nu} d^D \phi^{EF}_{i GH}$ & ($27,2,1/2$)  & $\Cap$ \tabularnewline 
\hline   
${C_{8,\overline{3},3,10}}^{BC}_{ ADEFG}$ & $G^{\mu\nu A}_B \overline{Q_L}^i_C \sigma_{\mu\nu} d^D \phi_i^{EFG}$ & ($10,2,1/2$)  & $\Cap$ \tabularnewline 
\hline  
${C_{8,\overline{3},3,\overline{10}}}^{BCEFG}_{ AD}$ & $G^{\mu\nu A}_B \overline{Q_L}^i_C \sigma_{\mu\nu} d^D \phi_{iEFG}$ & ($\overline{10},2,1/2$)  & $\Cap$ \tabularnewline 
\hline  
${C_{8,\overline{3},3,8}}^{BCF}_{ ADE}$ & $G^{\mu\nu A}_B \overline{Q_L}^i_C \sigma_{\mu\nu} d^D \phi_{iF}^{E}$ & ($8,2,1/2$)  & $\Cap$ \tabularnewline 
\hline  
${C_{8,\overline{3},3}}^{BC}_{ AD}$ & $G^{\mu\nu A}_B \overline{Q_L}^i_C \sigma_{\mu\nu} d^D \phi_{i}$ & ($1,2,1/2$)  & $\Cap$ \tabularnewline 
\hline 
${C_{8,3,3,\overline{24}}}^{BFGH}_{ ACDE}$ & $G^{\mu\nu A}_B \overline{u^c}^C \sigma_{\mu\nu} u^D \phi^E_{FGH}$ & ($\overline{24},1,-4/3$) & $\Cap$  \tabularnewline 
\hline  
${C_{8,3,3,15}}^{BG}_{ACDEF}$ & $G^{\mu\nu A}_B \overline{u^c}^C \sigma_{\mu\nu} u^D \phi^{EF}_G$ & (${15},1,-4/3$)  & $\Cap$ \tabularnewline 
\hline  
${C_{8,3,3,\overline{6}}}^{BEF}_{ACD}$  & $G^{\mu\nu A}_B \overline{u^c}^C \sigma_{\mu\nu} u^D \phi_{EF}$ & ($\overline{6},1,-4/3$) & $\Cap$  \tabularnewline 
\hline  
${C_{8,3,3,3}}^{B}_{ACDE}$ & $G^{\mu\nu A}_B \overline{u^c}^C \sigma_{\mu\nu} u^D \phi^E$ & (${3},1,-4/3$)  & $\Cap$ \tabularnewline 
\hline  
${C_{8,3,3,\overline{24}}}^{BFGH}_{ ACDE}$ & $G^{\mu\nu A}_B \overline{u^c}^C \sigma_{\mu\nu} d^D \phi^E_{FGH}$ & ($\overline{24},1,-1/3$) & $\Cap$  \tabularnewline 
\hline  
${C_{8,3,3,15}}^{BG}_{ACDEF}$ & $G^{\mu\nu A}_B \overline{u^c}^C \sigma_{\mu\nu} d^D \phi^{EF}_G$ & (${15},1,-1/3$)  & $\Cap$ \tabularnewline 
\hline  
${C_{8,3,3,\overline{6}}}^{BEF}_{ACD}$  & $G^{\mu\nu A}_B \overline{u^c}^C \sigma_{\mu\nu} d^D \phi_{EF}$ & ($\overline{6},1,-1/3$) & $\Cap$  \tabularnewline 
\hline  
${C_{8,3,3,3}}^{B}_{ACDE}$ & $G^{\mu\nu A}_B \overline{u^c}^C \sigma_{\mu\nu} d^D \phi^E$ & (${3},1,-1/3$)  & $\Cap$ \tabularnewline 
\hline  
${C_{8,3,3,\overline{24}}}^{BFGH}_{ ACDE}$ & $G^{\mu\nu A}_B \overline{d^c}^C \sigma_{\mu\nu} d^D \phi^E_{FGH}$ & ($\overline{24},1,2/3$) & $\Cap$  \tabularnewline 
\hline  
${C_{8,3,3,15}}^{BG}_{ACDEF}$ & $G^{\mu\nu A}_B \overline{d^c}^C \sigma_{\mu\nu} d^D \phi^{EF}_G$ & (${15},1,2/3$)  & $\Cap$ \tabularnewline 
\hline  
${C_{8,3,3,\overline{6}}}^{BEF}_{ACD}$  & $G^{\mu\nu A}_B \overline{d^c}^C \sigma_{\mu\nu} d^D \phi_{EF}$ & ($\overline{6},1,2/3$) & $\Cap$  \tabularnewline 
\hline  
${C_{8,3,3,3}}^{B}_{ACDE}$ & $G^{\mu\nu A}_B \overline{d^c}^C \sigma_{\mu\nu} d^D \phi^E$ & (${3},1,2/3$)  & $\Cap$ \tabularnewline 
\hline  
\end{tabular}
\end{center}
\caption{\label{tab:4}Dimension 6 operators of type $Gff\phi$, where both of the SM fermions are quarks and and $\phi$ is a scalar state.}
\end{table}

Within Table~\ref{tab:3}, which contains operators where at least one of the SM fermions is not a quark, we again have a range of quantum numbers. Operators containing the SU(3)$_\text{c}$ field strength tensor and no quarks include LEX states that are color octets. Conversely, operators with the SU(3)$_\text{c}$ field strength tensor and one quark follow the SU(3)$_\text{c}$ product rule 
\begin{equation}\label{eq:3}
    8\otimes 3 = 15 \oplus \overline{6} \oplus 3.
\end{equation}
We note that the 15-plet found in this product rule is the type with 2 upper and 1 lower fundamental SU(3) index ($\phi^{AB}_{C}$). 

\begin{figure}[h!] 
    \centering
    \includegraphics[width=0.3\linewidth]{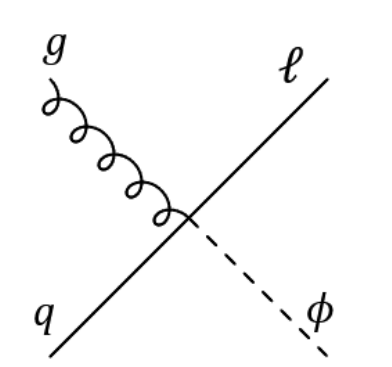} 
    \caption{Interaction vertex attainable through operators found in Table~\ref{tab:3}.}
    \label{fig:5}
\end{figure}

Operators found in Table~\ref{tab:3} lead to some interesting production and decay processes for LEX states. For example, operators involving triplet, sextet, or 15-plet LEX states couple a bosonic LEX state to a quark, gluon, and lepton. This vertex is shown in Fig.~\ref{fig:5}. This interaction allows for the associated production of a LEX particle with a SM lepton through quark-gluon fusion, and the decay of the LEX state into a 2-jet plus lepton resonance.

Finally, for the operators in Table~\ref{tab:4} which contain two quarks and a strong field strength tensor, we need the SU(3) tensor product rules (computed by us and cross verified with \cite{Slansky:1981yr})
\begin{equation}\label{eq:19}
8\otimes3\otimes3 = 8\otimes (6 \oplus \overline{3}) = 24\oplus \overline{15 } \oplus 6 \oplus \overline{3} \oplus \overline{15} \oplus 6 \oplus \overline{3}
\end{equation}
and
\begin{equation}\label{eq:20}
    8\otimes3\otimes \overline{3} = 8\otimes (8\oplus 1) = 27 \oplus \overline{10} \oplus 10 \oplus 8 \oplus 8 \oplus 1 \oplus 8.
\end{equation}
Operators with a quark and a charge conjugated antiquark will follow the algebra of Eq.~\ref{eq:19}, while operators with a quark and an antiquark will follow the algebra of Eq.~\ref{eq:20}.  We see that these operators allow us to access extremely exotic SU(3)$_\text{c}$ charges. The $\overline{24}$ of SU(3)$_\text{c}$ is an object with one upper and three lower fundamental indices, $\phi^A_{BCD}$.

The SU(2)$_\text{L}$ charge of the LEX state in operators of type $Gff\phi$ depends entirely on the left- or right-handedness of the SM fermion fields. If an operator of this type contains two right-handed SM fermions, all SM fields within the operator will be electroweak singlets; as such, the LEX state will also be a singlet. However, in the case that there is one left-handed and one right-handed SM fermion, the LEX state will be an SU(2)$_\text{L}$ doublet. Finally, in the case of two left-handed SM fermions, we look to the SU(2)$_\text{L}$ product rule $2\otimes 2 = 3 \oplus 1$; the LEX state will be a triplet or singlet.

We can thus singly produce, for example, a LEX 24-plet in association with a jet from quark-quark or quark-gluon fusion. This LEX state may have unit charge if it is a doublet or triplet of SU(2)$_\text{L}$. This opens the possibility of production signatures with electroweak cascade decays of the LEX state, depending on multiplet splitting and operator coefficients.  We can also see that the lightest state in the LEX multiplet is guaranteed a decay into three jets.  The phenomenology can be further enriched by allowing flavor texture of the operators. This might allow three body decays of LEX states to involve third generation quarks.

\subsubsection{Higher spin LEX $Gff\phi$ operators} \label{s3.4.1}
We now present operator structures for higher spin LEX states for $Gff \phi$  type operators. 
Working out the Lorentz representations for LEX states of this type, we have the tensor product
\begin{align}
\begin{split}
[(1,0)\oplus(0,1)] \otimes [(1/2,0)&\oplus (0,1/2)] \otimes [(1/2,0)\oplus (0,1/2)] \\
 &= [(2,0)\oplus(0,2)]\oplus3\cdot[(1,0)\oplus(0,1)] \oplus 2\cdot(0,0) \oplus 2\cdot(1,1) \\
 & \oplus 2\cdot [(3/2,1/2)\oplus (1/2,3/2)] \oplus4\cdot (1/2,1/2) .
\end{split}
\end{align}
The above expression contains 96 degrees of freedom. We thus find that the following operator structures are allowed for each SU(3) color representation of the LEX state. 

\begin{itemize}
\item $[(2,0)\oplus(0,2)]$:
Operators coupling to this LEX field must have the form
\begin{equation}
G_{\mu\nu}\overline{f}\sigma_{\rho\sigma}f\phi^{\mu\nu\rho\sigma} .
\end{equation}

\item $[(1,0)\oplus(0,1)]$:
We find operators with two possible Lorentz structures. These are 
\begin{equation}
G_{\mu\nu}\overline{f}f\phi^{\mu\nu} \quad \text{and} \quad  G_{\mu\nu}\overline{f}\sigma^{\nu}_{\rho} f\phi^{\mu\rho} .
\end{equation}

\item $(0,0)$: Operators with this scalar LEX field have been shown in Table~\ref{tab:3} and Table~\ref{tab:4}.

\item $(1,1)$: The Lorentz structure for operators containing this LEX state is
\begin{equation}
G_{\mu\nu}\overline{f}\sigma^{\nu}_{\rho} f\phi^{\mu\rho} .
\end{equation}

\item $(1/2,1/2)$:
Operators involving this LEX state have the form
\begin{equation}
G_{\mu\nu}\overline{f}\gamma^{\mu} f\phi^{\nu} .
\end{equation}

\item $[(3/2,1/2) \oplus(1/2,3/2)]$: 
Operators containing this LEX state have the Lorentz structures
\begin{equation}
G_{\mu\nu}\overline{f}\gamma_{\rho}f\phi^{\rho[\mu\nu]} \quad \text{and} \quad G_{\mu\nu}\overline{f}\gamma_{\rho}f\phi^{\mu[\rho\nu]} .
\end{equation}
\end{itemize}

\subsection{Operators of form $GDD\phi$ and $GDD\phi H$}
The next option for operators involving a scalar LEX field in the gluon-gluon portal  contains the strong field strength tensor, a LEX scalar, and two derivatives; these operators are dimension five. One can also add an additional Higgs boson in the operator at dimension six. In operators of either type, the derivatives can each act on any of the fields present. Below, we show that operators of this form are redundant to the portal for scalar LEX states.

In operators without the Higgs field, we can write three operators, which are all related by integration by parts:
\begin{align}
    D_\mu D_\nu G^{\mu\nu} \phi = D_\nu G^{\mu\nu} D_\mu \phi =  G^{\mu\nu} D_\nu D_\mu \phi.
\end{align}
As such, we can choose to keep only one of these options; here, we choose to work with $G^{\mu\nu} D_\nu D_\mu \phi$. We can use the identity
\begin{equation}\label{eq:7}
D^{\mu}D^{\nu}=\frac{{1}}{2}[D^{\mu},D^{\nu}] +\frac{{1}}{2}\{D^{\mu},D^{\nu}\}   
\end{equation}
in conjunction with the antisymmetric nature of the field strength tensor to rewrite this operator:
\begin{equation}\label{eq:8}
    G^{\mu\nu} D_\nu D_\mu \phi = \frac{1}{2} G^{\mu\nu} \left[ D_\nu, D_\mu \right] \phi.
\end{equation}
Then, we can use the relation
\begin{equation}\label{eq:9}
    [D^\mu, D^\nu] = -i eF^{\mu\nu}
\end{equation}
to show that this operator is redundant with those of form $GG\phi$. As such, we can drop operators of this type from further consideration in the $g$-$g$ portal.

A very similar approach can be taken for operators of type $GDD\phi H$. Integration by parts can reduce the number of operators from six to three. Then, again using the antisymmetric nature of the SU(3)$_\text{c}$ field strength tensor, along with Eq.~\ref{eq:7} and Eq.~\ref{eq:8}, it can be shown that these operators are redundant with those of type $GFH\phi$. We thus discard all operators of this type that involve a scalar LEX field.

\subsubsection{ Higher spin LEX states in Operators the form $GDD\phi$ $GDD\phi H$}
We now enumerate Lorentz structures for operators of the form $GDD\phi$. The Lorentz tensor product for LEX representations allowed within this operator is 
\begin{align}
\begin{split}
[(1,0)\oplus(0,1)]&\otimes(1/2,1/2) \otimes(1/2,1/2)  \\
&=[(2,1)\oplus(1,2)]\oplus 4 \cdot(1,1) \oplus 3\cdot [(1,0)\oplus(0,1)] \oplus 2\cdot (0,0) ,
\end{split}
\end{align}
where we have checked that both sides of the equation have 96 degrees of freedom. We find the following Lorentz structures for our operators, depending on the Lorentz structure of the LEX state.
\begin{itemize}
    \item $[(2,1)\oplus(1,2)]$: We find operators with the Lorentz structure
    \begin{equation}
    G_{\alpha\beta}D_{\mu}D_{\nu}\phi^{ \mu\nu [\alpha \beta]} .
    \end{equation}
    \item  $[(1,0)\oplus(0,1)]$: There are two relevant Lorentz structures. These are
    \begin{equation}
G_{\mu\nu}D^{\alpha}D_{\alpha}\phi^{ \mu\nu} \quad \text{and} \quad G_{\mu\alpha}D^{\alpha}D_{\beta}\phi^{ \beta\mu} .
\end{equation}
\item $(1,1)$: Due to the symmetry of the two indices within this spin 2 object, we have the Lorentz structure
\begin{equation}
G_{\mu\alpha}D^{\alpha}D_{\beta}\phi^{ \beta\mu},
\end{equation}
with other contractions forbidden due to the antisymmetry of the field strength tensor.
\end{itemize}


\subsection{Operators of form $Dff\phi$ and $Dff\phi H$}\label{s3.5}
We next consider operators which include two SM fermions, a scalar LEX state, and one derivative. We include operators with zero or one Higgs fields. Operators of this type can be included in the $g$-$q$ portal. Specifically, when a covariant derivative acts on a state charged under SU(3)$_\text{c}$, a gluon can be induced through the relation
\begin{equation}
D^{\mu}\Phi \supset \partial^{\mu} \Phi + (i g_3 \tau_3  A_3^{\mu} +i g_2 \tau_2 A_2^{\mu}  + ig_1 Y_{\phi} A_1^{\mu})\Phi,
\end{equation}
where $\Phi$ is an (SU(3)$_\text{c}$, SU(2)$_\text{L}$, U(1)$_Y$) charged state,  $A_3$ is the
gluon field, and $A_{2}$ is the SU(2)$_\text{L}$ gauge field. As such, each derivative acting on a field that is charged under SU(3)$_\text{c}$ can lead to interactions involving a gluon. However, operators of this type will not be part of the $g$-$g$ portal, as there is only one derivative present (and as such only one gluon can be generated). We note that, as part of the gluon-quark portal, at least one of the SM fermions must be a quark. The other SM fermion may be a quark or a lepton.

For operators of type $Dff\phi$, we have three potential operators and one integration by parts relation. This leaves us with two independent operators. In the case where any of the fields is a singlet under SU(3)$_\text{c}$, this number further decreases to one independent operator in the portal.  Operators of type $Dff\phi$ and $Dff\phi H $, which are dimension five and six respectively, are shown in Table~\ref{tab:5} and Table~\ref{tab:6} for scalar LEX states. Contraction of color indices should be identical between these two tables, so they are only explicitly written out in Table~\ref{tab:5}.
\begin{table}[h!]
\begin{center}
\begin{tabular}{|c|c|c|c|}
\hline 
CG Coefficients & Operators & (SU(3)$_{\text{c}}$, SU(2)$_{\text{L}}$, U(1)$_Y$) & Portal\tabularnewline
\hline 
\hline 
${C_{8,\overline{3},3}}^{AD}_{BC}$ & $D_\mu \overline{Q_L}^i_A \gamma^\mu  Q^B_{Lj} \phi^{jC}_{iD}, \quad  \overline{Q_L}^i_A  \gamma^\mu D_\mu Q^B_{Lj} \phi^{jC}_{iD}$ & ($8,3,0$) & $\star$ \tabularnewline 
\hline 
${C_{8,\overline{3},3}}^{AD}_{BC}$  &  $D_\mu \overline{Q_L}^i_A \gamma^\mu  Q^B_{Li} \phi^C_D, \quad  \overline{Q_L}^i_A  \gamma^\mu D_\mu Q_{Li}^B \phi^C_D$ & ($8,1,0$) & $\star$ \tabularnewline 
\hline  
${C_{\overline{3},3}}^A_B$ &  $D_\mu \overline{Q_L}_A^i \gamma^\mu  Q_{Lj}^B \phi^j_i, \quad  \overline{Q_L}_A^i  \gamma^\mu D_\mu Q^B_{Lj} \phi^j_i$ & ($1,3,0$) & $\star$ \tabularnewline 
\hline  
${C_{\overline{3},3}}^A_B$ &  $D_\mu \overline{Q_L}_A^i \gamma^\mu  Q_{Li}^B \phi, \quad  \overline{Q_L}_A^i  \gamma^\mu D_\mu Q_{Li}^B \phi$ & ($1,1,0$) & $\star$ \tabularnewline 
\hline  
${C_{\overline{6},3,3}}^{CD}_{AB}$ &  $D_\mu \overline{Q_L^c}^A_i \gamma^\mu  u^B \phi^i_{CD}, \quad  \overline{Q^c_L}^A_i  \gamma^\mu D_\mu u^B \phi^i_{CD}$ & ($\overline{6},2,-5/6$) & $\star$ \tabularnewline 
\hline  
${C_{3,3,3}}_{ABC}$  & $D_\mu \overline{Q_L^c}^A_i \gamma^\mu  u^B \phi^{iC}, \quad  \overline{Q^c_L}^A_i  \gamma^\mu D_\mu u^B \phi^{iC}$ & ($3,2,-5/6$) & $\star$ \tabularnewline 
\hline  
${C_{\overline{6},3,3}}^{CD}_{AB}$ &  $D_\mu \overline{Q_L^c}_i^A \gamma^\mu  d^B \phi^i_{CD}, \quad  \overline{Q^c_L}_i^A  \gamma^\mu D_\mu d^B \phi^i_{CD}$ & ($\overline{6},2,1/6$) & $\star$ \tabularnewline 
\hline 
${C_{3,3,3}}_{ABC}$ &  $D_\mu \overline{Q_L^c}_i^A \gamma^\mu  d^B \phi^{iC}, \quad  \overline{Q^c_L}^A_i  \gamma^\mu D_\mu d^B \phi^{iC}$ & ($3,2,1/6$) & $\star$ \tabularnewline 
\hline 
${C_{\overline{3},3}}^A_B$ & $D_\mu \overline{Q_L}^i_A \gamma^\mu  L_j \phi^{jB}_i$ & ($3,3,2/3$) & $\star$ \tabularnewline 
\hline  
${C_{\overline{3},3}}^A_B$ &  $D_\mu \overline{Q_L}_A^i \gamma^\mu  L_i \phi^B$ & ($3,1,2/3$) & $\star$ \tabularnewline 
\hline  
${C_{\overline{3},3}}^B_A$ & $D_\mu \overline{Q_L^c}_{i}^A \gamma^\mu  \ell \phi^{i}_B$ & ($\overline{3},2,5/6$) & $\star$ \tabularnewline 
\hline  
${C_{8,\overline{3},3}}^{AD}_{BC}$ & $D_\mu \overline{u}_A \gamma^\mu  u^B \phi^C_D, \quad  \overline{u}_A  \gamma^\mu D_\mu u^B \phi^C_D$ & ($8,1,0$) & $\star$ \tabularnewline 
\hline  
${C_{\overline{3},3}}^A_B$ & $D_\mu \overline{u}_A \gamma^\mu  u^B \phi, \quad  \overline{u}  \gamma^\mu D_\mu u \phi$ & ($1,1,0$) & $\star$ \tabularnewline 
\hline  
${C_{8,\overline{3},3}}^{AD}_{BC}$ & $D_\mu \overline{u}_A \gamma^\mu  d^B \phi^C_D, \quad  \overline{u}_A  \gamma^\mu D_\mu d^B \phi^C_D$ & ($8,1,1$) & $\star$ \tabularnewline 
\hline  
${C_{\overline{3},3}}^A_B$ & $D_\mu \overline{u}_A \gamma^\mu  d^B \phi, \quad  \overline{u}_A  \gamma^\mu D_\mu d^B \phi$ & ($1,1,1$) & $\star$ \tabularnewline 
\hline  
${C_{\overline{3},3}}_A^B$ & $D_\mu \overline{u^c}^A \gamma^\mu  L_i \phi^i_B$ & ($\overline{3},2,-1/6$) & $\star$ \tabularnewline 
\hline  
${C_{\overline{3},3}}^A_B$ & $D_\mu \overline{u}_A \gamma^\mu  \ell \phi^B$ & ($3,1,5/3$) & $\star$ \tabularnewline 
\hline  
${C_{8,\overline{3},3}}^{AD}_{BC}$ & $D_\mu \overline{d}_A \gamma^\mu  d^B \phi^C_D, \quad  \overline{d}_A  \gamma^\mu D_\mu d^B \phi^C_D$ & ($8,1,0$) & $\star$ \tabularnewline 
\hline 
${C_{\overline{3},3}}^A_B$ & $D_\mu \overline{d}_A \gamma^\mu  d^B \phi, \quad  \overline{d}_A  \gamma^\mu D_\mu d^B \phi$ & ($1,1,0$) & $\star$ \tabularnewline 
\hline 
${C_{\overline{3},3}}_A^B$ & $D_\mu \overline{d^c}^A \gamma^\mu  L_i \phi^i_B$ & ($\overline{3},2,5/6$) & $\star$ \tabularnewline 
\hline  
${C_{\overline{3},3}}^A_B$ & $D_\mu \overline{d}_A \gamma^\mu  \ell \phi^B$ & ($3,1,2/3$) & $\star$ \tabularnewline 
\hline  
\end{tabular}
\end{center}
\caption{\label{tab:5}Dimension 5 operators of type $Dff\phi$.}
\end{table}

\begin{table}[h!]
\begin{center}
\begin{tabular}{|c|c|c|}
\hline 
Operators & (SU(3)$_{\text{c}}$, SU(2)$_{\text{L}}$, U(1)$_Y$) & Portal\tabularnewline
\hline 
\hline 
$D_\mu \overline{Q_L}^i \gamma^\mu  Q_{Lj} \phi_i^{jk} H_k, \quad  \overline{Q_L}^i  \gamma^\mu D_\mu Q_{Lj} \phi_i^{jk} H_k$ & ($8,4,-1/2$),  ($1,4,-1/2)$ & $\star$ \tabularnewline 
\hline  
$D_\mu \overline{Q_L}^i \gamma^\mu  Q_{Lj} \phi^j H_i, \quad  \overline{Q_L}^i  \gamma^\mu D_\mu Q_{Lj} \phi^j H_i$ & \multirow{2}{*}{($8,2,-1/2$), ($1,2,-1/2)$} & \multirow{2}{*}{$\star$} \tabularnewline 
$D_\mu \overline{Q_L}^i \gamma^\mu  Q_{Li} \phi^j H_i, \quad  \overline{Q_L}^i  \gamma^\mu D_\mu Q_{Li} \phi^j H_j$ & & \tabularnewline
\hline  
$D_\mu \overline{Q_L^c}_i \gamma^\mu  u \phi^{ij} H_j, \quad  \overline{Q^c_L}_i  \gamma^\mu D_\mu u \phi^{ij} H_j$ & ($\overline{6},3,-4/3$), ($3,3,-4/3$) & $\star$ \tabularnewline 
\hline  
$D_\mu \overline{Q_L^c}_i \gamma^\mu  u \phi^{i}_j H^{\dagger j}, \quad  \overline{Q^c_L}_i  \gamma^\mu D_\mu u \phi^{i}_j H^{\dagger j}$ & ($\overline{6},3,-1/3$), ($3,3,-1/3$) & $\star$ \tabularnewline 
\hline  
$D_\mu \overline{Q_L^c}^i \gamma^\mu  u \phi H_i, \quad  \overline{Q^c_L}^i  \gamma^\mu D_\mu u \phi H_i$ & ($\overline{6},1,-4/3$), ($3,1,-4/3$) & $\star$ \tabularnewline 
\hline  
$D_\mu \overline{Q_L^c}_i \gamma^\mu  u \phi H^{\dagger i}, \quad  \overline{Q^c_L}_i  \gamma^\mu D_\mu u \phi H^{\dagger i}$ & ($\overline{6},1,-1/3$), ($3,1,-1/3$) & $\star$ \tabularnewline 
\hline  
$D_\mu \overline{Q_L^c}_i \gamma^\mu  d \phi^{ij} H_j, \quad  \overline{Q^c_L}_i  \gamma^\mu D_\mu d \phi^{ij} H_j$ & ($\overline{6},3,-1/3$), ($3,3,-1/3$) & $\star$ \tabularnewline 
\hline  
$D_\mu \overline{Q_L^c}_i \gamma^\mu  d \phi^{i}_j H^{\dagger j}, \quad  \overline{Q^c_L}_i  \gamma^\mu D_\mu d \phi^{i}_j H^{\dagger j}$ & ($\overline{6},3,2/3$), ($3,3,2/3$) & $\star$ \tabularnewline 
\hline  
$D_\mu \overline{Q_L^c}^i \gamma^\mu  d \phi H_i, \quad  \overline{Q^c_L} ^i \gamma^\mu D_\mu d \phi H_i$ & ($\overline{6},1,-1/3$), ($3,1,-1/3$) & $\star$ \tabularnewline 
\hline  
$D_\mu \overline{Q_L^c}_i \gamma^\mu  d \phi H^{\dagger i}, \quad  \overline{Q^c_L}_i \gamma^\mu D_\mu d \phi H^{\dagger i}$ & ($\overline{6},1,2/3$), ($3,1,2/3$) & $\star$ \tabularnewline 
\hline  
$D_\mu \overline{Q_L}^i \gamma^\mu  L_j \phi^{jk}_i H_k$ & ($3,4,1/6$) & $\star$ \tabularnewline 
\hline 
$D_\mu \overline{Q_L}^i \gamma^\mu  L_j \phi^{j}_{ik} H^{\dagger k}$ & ($3,4,7/6$) & $\star$ \tabularnewline 
\hline 
$D_\mu \overline{Q_L}^i \gamma^\mu  L_i \phi^j H_j$, $D_\mu \overline{Q_L}^i \gamma^\mu  L_j \phi^j H_i$ & ($3,2,1/6$) & $\star$ \tabularnewline 
\hline  
$D_\mu \overline{Q_L}^i \gamma^\mu  L_i \phi_j H^{\dagger j}$, $D_\mu \overline{Q_L}_i \gamma^\mu  L_j \phi^j H^{\dagger i}$ & ($3,2,7/6$) & $\star$ \tabularnewline 
\hline  
$D_\mu \overline{Q_L^c}_i \gamma^\mu  \ell \phi^{ij} H_j$ & ($\overline{3},3,1/3$) & $\star$ \tabularnewline 
\hline  
$D_\mu \overline{Q_L^c}_i \gamma^\mu  \ell \phi^{i}_j H^{\dagger j}$ & ($\overline{3},3,4/3$) & $\star$ \tabularnewline 
\hline  
$D_\mu \overline{Q_L^c}^i \gamma^\mu  \ell \phi H_i$ & ($\overline{3},1,1/3$) & $\star$ \tabularnewline 
\hline  
$D_\mu \overline{Q_L^c}_i \gamma^\mu  \ell \phi H^{\dagger i}$ & ($\overline{3},1,4/3$) & $\star$ \tabularnewline 
\hline  
$D_\mu \overline{u} \gamma^\mu  u \phi^i H_i, \quad  \overline{u}  \gamma^\mu D_\mu u \phi^i H_i$ & ($8,2,-1/2$), ($1,2,-1/2$) & $\star$ \tabularnewline 
\hline  
$D_\mu \overline{u} \gamma^\mu  d \phi^i H_i, \quad  \overline{u}  \gamma^\mu D_\mu d \phi^i H_i$ & ($8,2,1/2$), ($1,2,1/2$) & $\star$ \tabularnewline 
\hline  
$D_\mu \overline{u} \gamma^\mu  d \phi_i H^{\dagger i}, \quad  \overline{u}  \gamma^\mu D_\mu d \phi_i H^{\dagger i}$ & ($8,2,3/2$), ($1,2,3/2$) & $\star$ \tabularnewline 
\hline  
$D_\mu \overline{u^c} \gamma^\mu  L_i \phi^{ij} H_j$ & ($\overline{3},3,-2/3$) & $\star$ \tabularnewline 
\hline  
$D_\mu \overline{u^c} \gamma^\mu  L_i \phi^{i}_j H^{\dagger j}$ & ($\overline{3},3,1/3$) & $\star$ \tabularnewline 
\hline  
$D_\mu \overline{u^c} \gamma^\mu  L^i \phi^{} H_i$ & ($\overline{3},1,-2/3$) & $\star$ \tabularnewline 
\hline  
$D_\mu \overline{u^c} \gamma^\mu  L_i \phi^{} H^{\dagger i}$ & ($\overline{3},1,1/3$) & $\star$ \tabularnewline 
\hline  
$D_\mu \overline{u} \gamma^\mu  \ell \phi^i H_i$ & ($3,2,7/6$) & $\star$ \tabularnewline 
\hline 
$D_\mu \overline{u} \gamma^\mu  \ell \phi_i H^{\dagger i}$ & ($3,2,13/6$) & $\star$ \tabularnewline 
\hline 
$D_\mu \overline{d} \gamma^\mu  d \phi^i H_i, \quad  \overline{d}  \gamma^\mu D_\mu d \phi^i H_i$ & ($8,2,-1/2$), ($1,2,-1/2$) & $\star$ \tabularnewline 
\hline  
$D_\mu \overline{d^c} \gamma^\mu  L_i \phi^{ij} H_j$ & ($\overline{3},3,1/3$) & $\star$ \tabularnewline 
\hline  
$D_\mu \overline{d^c} \gamma^\mu  L_i \phi^{i}_j H^{\dagger j}$ & ($\overline{3},3,4/6$) & $\star$ \tabularnewline 
\hline  
$D_\mu \overline{d^c} \gamma^\mu  L_i \phi H^{\dagger i}$ & ($\overline{3},1,4/6$) & $\star$ \tabularnewline 
\hline  
$D_\mu \overline{d} \gamma^\mu  \ell \phi^i H^{}_i$ & ($3,2,1/6$) & $\star$ \tabularnewline 
\hline 
$D_\mu \overline{d} \gamma^\mu  \ell \phi_i H^{\dagger i}$ & ($3,2,7/6$) & $\star$ \tabularnewline 
\hline 
\end{tabular}
\end{center}
\caption{\label{tab:6}Dimension 6 operators of type $Dff\phi H $ for scalar LEX states. SU(3)$_\text{c}$ indices are not written out.}
\end{table}

For operators of type $Dff\phi$ and $Dff\phi H$, the SU(3)$_\text{c}$ charge of the LEX state is dependent on the quark content within each operator. For operators with only one quark, the LEX state is a color (anti)triplet. Operators containing two quarks follow either the SU(3)$_\text{c}$ product rule
\begin{equation}\label{eq:11}
3\otimes 3 = \overline{3} \oplus 6
\end{equation}
or 
\begin{equation}\label{eq:12}
    3\otimes \overline{3} = 1\oplus 8.
\end{equation}
Meanwhile, the handedness of the SM fermions and the presence of a Higgs boson determine the electroweak charge of the LEX state. In the operators with no Higgs present, the LEX SU(2)$_\text{L}$ charges follow the same rules as in operators of form $Gff\phi$, discussed in Section~\ref{s3.3}. However, the operators containing a Higgs state require additional algebra. In the simplest case, of operators containing two right-handed SM fermions, the LEX state will be an SU(2)$_\text{L}$ doublet. In operators with one left-handed SM fermion, the algebra $2\otimes 2 = 3\oplus 1$ still tells us that the LEX state will be an electroweak triplet or singlet. Finally, in operators containing both a Higgs boson and two left-handed SM fermions, we require the SU(2)$_\text{L}$ algebra
\begin{equation}
    2\otimes 2 \otimes 2 = (3\oplus 1) \otimes 2 = 4\oplus 2 \oplus 2 .
\end{equation}
In this case, the LEX scalar will be an electroweak quadruplet or doublet.

\subsubsection{Higher spin operators of the form $Dff\phi, Dff\phi H$}
We now work out Lorentz structures for higher spin LEX states of the form $Dff\phi$ and $Dff\phi H$. The derivatives are in the Lorentz representation $(1/2,1/2)$, while SM Dirac fermions are in $[(1/2,0)\oplus(0,1/2)]$. Therefore, the Lorentz tensor product we need is 
\begin{align}
\begin{split}
[(1/2,0)&\oplus(0,1/2)]\otimes[(1/2,0)\oplus(0,1/2)]\otimes(1/2,1/2)  \\
&=[(3/2,1/2)\oplus(1/2,3/2)]\oplus 4\cdot(1/2,1/2)\oplus 2 \{ [(1,0)\oplus(0,1)]\oplus(1,1)\oplus(0,0) \} .
\end{split}
\end{align}
We can check that both terms above contain 64 degrees of freedom. The allowed Lorentz structures are listed for each of these spin states.
\begin{itemize}
\item $(0,0)$: Operators with this LEX state are shown in Table~\ref{tab:5} and \ref{tab:6}.
\item $(1,1)$: The Lorentz structure of the operator that includes this term is
\begin{equation}
D_{\mu}\overline{f}\gamma_{\nu}f\Phi^{\mu\nu} .
\end{equation}

\item $[(1,0)\oplus(0,1)]$: The relevant operator can (again) be written as
\begin{equation}
D_{\mu}\overline{f}\gamma_{\nu}f\Phi^{\mu\nu} .
\end{equation}

\item $[(3/2,1/2)\oplus(1/2,3/2)]$: The Lorentz structure of this operator is
\begin{equation}
D_{\mu}\overline{f}\sigma_{\sigma\rho}f\Phi^{\mu\sigma\rho} .
\end{equation}

\item $(1/2,1/2)$: An operator containing a LEX state in this representation has the structure
\begin{equation}
D_{\mu}\overline{f}f\Phi^{\mu} .
\end{equation}
\end{itemize}

\subsection{Operators of form $ffDD\phi$}\label{s2.5}
Up to dimension six, there is one final operator structure which must be considered for scalar LEX states. This operator type includes two SM fermions, two covariant derivatives, and a scalar LEX state. Operators of this type could be included in either the gluon-gluon or gluon-quark portal. At least one of the SM fermions must be a quark.

The $g$-$g$ portal, which requires two gluons, would necessitate the presence of two derivatives in the case where there is no SU(3)$_\text{c}$ field strength tensor. In order to introduce non-singlet SU(3)$_\text{c}$ charge, we require at least one of our SM fermions to be a quark. As such, operators of this type are candidates in both the $g$-$g$ and $g$-$q$ portals.
The two-derivative structure may look simple on first glance, but in reality it can take several different forms. Specifically, we must consider the forms
\begin{align}
    D_\mu D_\nu \sigma^{\mu\nu}, \qquad D_\mu D_\nu \gamma^{\mu}\gamma^{\nu}, \qquad D_\mu D^\mu.
\end{align}
We choose to start with the first of these forms, $D_\mu D_\nu \sigma^{\mu\nu}$. As we have operators with three fields and two derivatives, we may write six potential operators in the case that all fields present are charged under SU(3)$_\text{c}$:
\begin{align}
\begin{split}
    D_{\mu}D_{\nu}\overline{f_1}\sigma^{\mu\nu}f_2\phi, \qquad 
    \overline{f_1}\sigma^{\mu\nu} D_{\mu}D_{\nu}f_2\phi, \qquad 
    \overline{f_1}\sigma^{\mu\nu}f_2 D_{\mu}D_{\nu} \phi \\ 
    D_{\mu}\overline{f_1}\sigma^{\mu\nu}D_{\nu} f_2\phi, \qquad
    D_{\mu}\overline{f_1}\sigma^{\mu\nu} f_2D_{\nu} \phi, \qquad
    \overline{f_1}\sigma^{\mu\nu}D_{\mu} f_2 D_{\nu} \phi.
\end{split}    
\end{align}
However, integration by parts reduces our number of operators to three. We choose to continue working with the operators where both derivatives act on the same field. In this case, we again use Eq.~\ref{eq:7} and Eq.~\ref{eq:9}, in conjunction with the antisymmetry of $\sigma^{\mu\nu}$. This leaves us with operators of type $Fff\phi$, which are either redundant with those of Section~\ref{s3.3}, or which do not contribute to the gluon-gluon or gluon-quark portals. While we examined the case where all fields within our operators are charged under the strong force, the same arguments will work for the case where they are not. In this case, we would simply have fewer operators under consideration from the beginning.

The next operator structure that we will consider contains the derivative form $D_\mu D_\nu \gamma^{\mu}\gamma^{\nu}$. However, in this case we can insert the gamma matrix identity 
\begin{equation}
    \gamma^{\mu}\gamma^{\nu}=g^{\mu \nu}-i \sigma^{\mu\nu}.
\end{equation}
This identity shows us that operators containing the two-derivative Lorentz structure $D_\mu D_\nu \gamma^{\mu}\gamma^{\nu}$ are redundant with those containing $D_\mu D_\nu \sigma^{\mu\nu}$ and $D_\mu D^\mu$. We have already shown that the first of these forms is redundant within our portal, and we now choose to work with the latter, while considering the $g$-$g$ portal. 

Just as before, integration by parts reduces six operators with the derivative form  $D_\mu D^\mu$ to only three. In operators where one or more state is a singlet of SU(3)$_\text{c}$, the number of operators decreases further. Operators of this type are shown in Table~\ref{tab:7} for scalar LEX states within the gluon-gluon portal.
\begin{table}[h!]
\begin{center}
\begin{tabular}{|c|c|c|c|}
\hline 
CG Coefficients & Operators & (SU(3)$_{\text{c}}$, SU(2)$_{\text{L}}$, U(1)$_Y$) & Portal \tabularnewline
\hline 
\hline 
\multirow{2}{*}{${C_{\overline{6},3,3}}^{CD}_{AB}$} & $D^2 \overline{Q^c_L}_i^A  Q_{Lj}^B \phi^{ij}_{CD}, \quad  \overline{Q^c_L}^A_i D^2 Q_{Lj}^B \phi^{ij}_{CD}, $ & \multirow{2}{*}{($\overline{6},3,-1/3$)} & \multirow{2}{*}{$\diamond$}  \tabularnewline 
& $\overline{Q^c_L}^A_i  Q_{Lj}^B D^2\phi_{CD}^{ij}$ & & \tabularnewline
\hline  
\multirow{2}{*}{${C_{3,3,3}}^{}_{ABC}$} & $D^2 \overline{Q^c_L}^A_i  Q_{Lj}^B \phi^{ij C}, \quad  \overline{Q^c_L}^A_i D^2 Q^B_{Lj} \phi^{ij C}, $ & \multirow{2}{*}{($3,3,-1/3$)} & \multirow{2}{*}{$\diamond$}  \tabularnewline 
& $ \overline{Q^c_L}^A_i  Q_{Lj}^B D^2\phi^{ij C}$ & & \tabularnewline
\hline  
\multirow{2}{*}{${C_{\overline{6},3,3}}^{CD}_{AB}$} & $D^2 \overline{Q^c_L}^{iA}  Q_{Li}^B \phi_{CD}, \quad  \overline{Q^c_L}^{iA} D^2 Q_{Li}^B \phi_{CD},  $ & \multirow{2}{*}{($\overline{6},1,-1/3$)} & \multirow{2}{*}{$\diamond$}  \tabularnewline
& $\overline{Q^c_L}^{iA}  Q_{Li}^B D^2\phi_{CD}$ & & \tabularnewline
\hline  
\multirow{2}{*}{${C_{3,3,3}}^{}_{ABC}$} & $D^2 \overline{Q^c_L}^{iA}  Q_{Li}^B \phi^C, \quad  \overline{Q^c_L}^{iA} D^2 Q_{Li}^B \phi^C,  $ & \multirow{2}{*}{($3,1,-1/3$)} & \multirow{2}{*}{$\diamond$}  \tabularnewline 
& $\overline{Q^c_L}^{iA}  Q_{Li}^B D^2\phi^C$ & & \tabularnewline
\hline  
\multirow{2}{*}{${C_{8,\overline{3},3}}^{AD}_{BC}$} & $D^2 \overline{Q_L}^i_A u^B \phi^C_{iD} \quad \overline{Q_L}_A^i D^2 u^B \phi^C_{iD}, $ & \multirow{2}{*}{($8,2,-1/2$)} & \multirow{2}{*}{$\diamond$}  \tabularnewline
& $\overline{Q_L}_A^i u^B D^2 \phi^C_{iD}$ & & \tabularnewline
\hline
\multirow{2}{*}{${C_{\overline{3},3}}^{A}_{B}$} & $D^2 \overline{Q_L}^i_A u^B \phi_i \quad \overline{Q_L}^i_A D^2 u^B \phi_i,  $ & \multirow{2}{*}{($1,2,-1/2$)} & \multirow{2}{*}{$\diamond$}  \tabularnewline 
& $\overline{Q_L}_A^i u^B D^2 \phi_i$ & & \tabularnewline
\hline
\multirow{2}{*}{${C_{8,\overline{3},3}}^{AD}_{BC}$} & $D^2 \overline{Q_L}^i_A  d^B \phi^C_{iD}, \quad \overline{Q_L}_A^i  D^2 d^B \phi^C_{iD},  $ & \multirow{2}{*}{($8,2,1/2$)}  & \multirow{2}{*}{$\diamond$} \tabularnewline 
& $\overline{Q_L}^i_A  d^B D^2 \phi^C_{iD}$ & & \tabularnewline
\hline
\multirow{2}{*}{${C_{\overline{3},3}}^{A}_{B}$} & $D^2 \overline{Q_L}^i_A  d^B \phi_{i}, \quad \overline{Q_L}_A^i  D^2 d^B \phi_{i},  $ & \multirow{2}{*}{($1,2,1/2$)} & \multirow{2}{*}{$\diamond$}  \tabularnewline 
& $\overline{Q_L}^i_A  d^B D^2 \phi_i$ & & \tabularnewline
\hline
${C_{\overline{3},3}}^{B}_{A}$ & $D^2 \overline{Q^c_L}_i^A L_j \phi^{ij}_B, \quad \overline{Q^c_L}_i^A L_j D^2 \phi^{ij}_B$ & ($\overline{3},3,1/3$) & $\diamond$  \tabularnewline 
\hline  
${C_{\overline{3},3}}^{B}_{A}$ & $D^2 \overline{Q^c_L}^{iA} L_i \phi_B, \quad \overline{Q^c_L}^{iA} L_i D^2 \phi_B$ & ($\overline{3},1,1/3$) & $\diamond$  \tabularnewline 
\hline  
${C_{\overline{3},3}}^{A}_{B}$ & $D^2 \overline{Q_L}_A^i \ell \phi^B_i, \quad \overline{Q_L}_A^i \ell D^2 \phi^B_i$ & ($3,2,7/6$)  & $\diamond$ \tabularnewline 
\hline
\multirow{2}{*}{${C_{\overline{6},3,3}}^{CD}_{AB}$}  & $D^2 \overline{u^c}^A u^B \phi_{CD}, \quad \overline{u^c}^A D^2 u^B \phi_{CD}, $ & \multirow{2}{*}{($\overline{6},1,-4/3$)} & \multirow{2}{*}{$\diamond$}  \tabularnewline
& $\overline{u^c}^A u^B D^2\phi_{CD}$ & & \tabularnewline
\hline
\multirow{2}{*}{${C_{3,3,3}}^{}_{ABC}$} & $D^2 \overline{u^c}^A u^B \phi^C, \quad \overline{u^c}^A D^2 u^B \phi^C,  $ & \multirow{2}{*}{($3,1,-4/3$)} & \multirow{2}{*}{$\diamond$}  \tabularnewline 
& $\overline{u^c}^A u^B D^2\phi^C$ & & \tabularnewline
\hline
\multirow{2}{*}{${C_{\overline{6},3,3}}^{CD}_{AB}$}  & $D^2 \overline{u^c}^A  d^B \phi_{CD}, \quad \overline{u^c}^A  D^2d^B \phi_{CD}, $ & \multirow{2}{*}{($\overline{6},1,-1/3$)} & \multirow{2}{*}{$\diamond$}  \tabularnewline 
& $\overline{u^c}^A  d^B D^2\phi_{CD}$ & & \tabularnewline
\hline
\multirow{2}{*}{${C_{3,3,3}}^{}_{ABC}$} & $D^2 \overline{u^c}^A  d^B \phi^C, \quad \overline{u^c}^A  D^2d^B \phi^C, $ & \multirow{2}{*}{($3,1,-1/3$)} & \multirow{2}{*}{$\diamond$}  \tabularnewline 
& $ \overline{u^c}^A  d^B D^2\phi^C$ & & \tabularnewline
\hline
${C_{\overline{3},3}}^{A}_B$ & $D^2 \overline{u}_A L_i \phi^{iB}, \quad \overline{u}_A L_i D^2 \phi^{iB}$ & ($3,2,7/6$)  & $\diamond$ \tabularnewline 
\hline  
${C_{\overline{3},3}}^{B}_{A}$ & $D^2 \overline{u^c}^A  \ell \phi_B, \quad \overline{u^c}^A  \ell D^2 \phi_B$ & ($\overline{3},1,1/3$) & $\diamond$  \tabularnewline 
\hline
\multirow{2}{*}{${C_{\overline{6},3,3}}^{CD}_{AB}$}  & $D^2 \overline{d^c}^A d^B \phi_{CD}, \quad \overline{d^c}^A D^2 d^B \phi_{CD},  $ & \multirow{2}{*}{($\overline{6},1,2/3$)} & \multirow{2}{*}{$\diamond$}  \tabularnewline 
& $\overline{d^c}^A d^B D^2 \phi_{CD}$ & & \tabularnewline
\hline
\multirow{2}{*}{${C_{3,3,3}}^{}_{ABC}$} & $D^2 \overline{d^c}^A d^B \phi^C, \quad \overline{d^c}^A D^2 d^B \phi^C,  $ & \multirow{2}{*}{($3,1,2/3$)}  & \multirow{2}{*}{$\diamond$} \tabularnewline
& $\overline{d^c}^A d^B D^2 \phi^C$ & & \tabularnewline
\hline
${C_{\overline{3},3}}^{A}_B$ & $D^2 \overline{d}_A L_i \phi^{iB}, \quad \overline{d}_A L_i D^2\phi^{iB}$ & ($3,2,1/6$)  & $\diamond$ \tabularnewline 
\hline  
${C_{\overline{3},3}}^{B}_A$ & $D^2 \overline{d^c}^A \ell \phi_B, \quad \overline{d^c}^A \ell D^2 \phi_B$ & ($\overline{3},1,4/3$)  & $\diamond$ \tabularnewline 
\hline
\end{tabular}
\end{center}
\caption{\label{tab:7}Dimension 6 operators of type $ffDD\phi$ in the $g$-$g$ portal.}
\end{table}

Within Table~\ref{tab:7}, we find a range of quantum numbers. For operators with a quark and a charge conjugated anti-quark, we look to the SU(3)$_\text{c}$ product rule given in Eq.~\ref{eq:11}.
As such, the LEX state will be a triplet or sextet conjugate under SU(3)$_\text{c}$. Meanwhile, for operators with a quark and an anti-quark, we look to
Eq.~\ref{eq:12}---the LEX state will be a color octet or singlet. The SU(2)$_\text{L}$ charges will be assigned as in Section~\ref{s3.3}. Operators with two left-handed SM fermions contain LEX states that are electroweak triplets or singlets, while operators with one left-handed and one right-handed SM fermion will contain LEX states that are electroweak doublets. Finally, operators with two right-handed SM fermions will contain LEX states that are electroweak singlets.

At first glance, one would think that the operators shown in Table~\ref{tab:7} should also exist in the gluon-quark portal. However, we now continue to the $D_\mu D_\nu \gamma^{\mu}\gamma^{\nu}$ structure, instead of the $D_\mu D^\mu$ structure. Again, we know that there are six potential operators of this form, which can be reduced through three equations of motion. Following a similar pattern to Section 4.2.1 of~\cite{Carpenter:2024hvp}, we choose to keep the operators
\begin{equation}
    \slashed{D}\slashed{D}\overline{f_1}f_2\phi, \quad \slashed{D}\overline{f_1}\slashed{D}f_2\phi, \quad 
    \overline{f_1}\slashed{D}\slashed{D}f_2\phi .
\end{equation}
 Fermionic equations of motion then allow the remaining operators of this type to transform into those of type $Dff\phi H$, which was discussed for the gluon-quark portal in Section~\ref{s3.5}. All operators of this type are redundant within the $g$-$q$ portal.

 \subsubsection{$ffDD\phi$ operators with higher spin LEX states} 
 To compute the allowed LEX state Lorentz representations, we need the tensor product
 \begin{align}
 \begin{split}
 (1/2,1/2)&\otimes(1/2,1/2)\otimes[(1/2,0)\oplus(0,1/2)]\otimes[(1/2,0)\oplus(0,1/2)] \\
  &=[(2,1) \oplus (1,2)] \oplus [(2,0) \oplus (0,2)] \oplus 6\cdot(1,1) \oplus 5\cdot[(1,0) \oplus (0,1)] \oplus 4\cdot(0,0)\\
  &~~~~\oplus   2 \cdot (3/2,3/2) \oplus 4 \cdot [(3/2,1/1) \oplus (1/2,3/2)] \oplus 8 \cdot (1/2,1/2) .
 \end{split}
\end{align}
This yields 256 degrees of freedom.
Lorentz structures are as follows.
\begin{itemize}
\item (0,0): Operators involving LEX scalars have been enumerated in Table~\ref{tab:7}.

\item $[(1,0)\oplus(0,1)]$: As the LEX state in this representation has two antisymmetric indices, the allowed operator structures are
\begin{equation}
\overline{f}\sigma^{\mu\nu}f D_{\alpha}D^{\alpha}\phi_{\mu\nu} \quad \text{and} \quad \overline{f}\sigma^{\mu\nu}f D_{\nu}D^{\sigma}\phi_{\mu\sigma} .
\end{equation}

\item $(1/2,1/2)$: We have the allowed Lorentz structures
\begin{equation}
\overline{f}\gamma^{\mu}fD_{\alpha}D^{\alpha}\phi_{\mu} \quad \text{and} \quad \overline{f}\gamma^{\mu}fD^{\nu}D_{\mu}\phi_{\nu} .
\end{equation}

\item $(3/2,1/2)$ The only allowed Lorentz structure containing a LEX state with this Lorentz representation is 
\begin{equation}
\overline{f}\gamma^{\sigma}fD^{\nu}D^{\mu}\phi_{[\sigma \nu]\mu} .
\end{equation}

\item $(3/2,3/2)$:  We find the Lorentz structure
\begin{equation}
\overline{f}\gamma^{\sigma}fD^{\nu}D^{\mu}\Phi_{\sigma \nu\mu} .
\end{equation}

\item $(1,1)$: An operator containing a LEX state with this Lorentz representation can be written as
\begin{equation}
\overline{f}f D_{\nu}D_{\nu}\Phi^{\mu \nu} .
\end{equation}

\item $[(2,0)\oplus(0,2)]$: Due to symmetry considerations, the allowed Lorentz structure for this operator is 
\begin{equation}
\overline{f}\sigma^{\mu\nu}fD^{\sigma}D^{\rho}\phi_{[\mu\sigma] [\nu\rho]} .
\end{equation}

\item $[(2,1)\oplus(1,2)]$: Due to the symmetry considerations of this spin 3 LEX state, we must have an operator with the Lorentz structure
\begin{equation}
\overline{f}\sigma^{\mu\nu}fD^{\sigma}D^{\rho}\phi_{[\mu\nu] \{ \sigma\rho \}} .
\end{equation}
\end{itemize}

\subsection{Some Dimension 7 Operators: Operators of form $GFF\phi$}
Finally, we find it important to note a few of the most interesting operators at dimension seven. Our choice to limit this catalog to fully explore through dimension six was motivated by higher dimensions being increasingly suppressed by additional factors of the effective cutoff. However, this choice was somewhat arbitrary and the authors have previously engaged with  dimension seven exotic EFTs  with detectable collider signatures~\cite{Carpenter:2021gpl}. Constructing dimension seven operators allows us to access particles in spectacular SU(3)$_\text{c}$ representations and spins. Some of the most interesting types of dimension seven operators are those containing one LEX boson and three SM field strength tensors, or operators of type $GFF\phi$. These operators fall into only the $g$-$g$ portal. 

\subsubsection{Considering Lorentz Structures} \label{s3.8.1}
We will briefly note the Lorentz structures of possible operators of type $GFF\phi$. As SM field strength tensors are in the $(1,0)\oplus(0,1)$ Lorentz representation, we compute  the tensor product
\begin{align}
\begin{split}
&[(1,0)\oplus (0,1) ]\otimes [(1,0)\oplus (0,1) ] \otimes [(1,0)\oplus (0,1) ]  \\
&= [(3,0)\oplus(0,3)] \oplus 2[(2,0)\oplus(0,2)] \oplus 
6[(1,0)\oplus(0,1)]+ 2[(0,0)] \oplus 3[(2,1)\oplus(1,2)] \oplus 6(1,1) ,
\end{split}
\end{align}
where we can crosscheck that there are 216 degrees of freedom in the total tensor product.
From this, we see that we may access spin 3, 2, 1, and 0 bosons in operators of this type. The Lorentz structures of operators including the various spin states will now be listed.
\begin{itemize}
    \item $(0,0)$: For this LEX Lorentz structure, we have the operator form
    \begin{equation}
        F_{\mu\nu}F^{\nu}_{~\rho} F^{\rho \mu}\phi .
    \end{equation}
    \item $(1,0) \oplus(0,1)$: This spin 1 LEX state has the potential allowed Lorentz structures 
    \begin{equation}
        F_{\mu\nu}F^{\mu\nu} F_{\rho \sigma}\phi^{\rho\sigma} \quad \text{and} \quad F_{\mu\nu}F^{\mu\rho} F_{\rho \sigma}\phi^{\mu\sigma}.
    \end{equation}
    \item $(1,1)$: This symmetric object cannot be contracted with a totally antisymmetric $F^{\mu\nu}$. As such, we find the only allowed Lorentz structure to be 
    \begin{equation}
    F_{\mu\nu}F^{\mu\rho} F_{\rho \sigma}\Phi^{ \{ \nu\sigma\}}.
    \end{equation}
    \item $(2,0)\oplus (0,2)$: We find the Lorentz structure 
    \begin{equation}
        F^{\mu\nu}F^{\rho}_{\alpha}F^{\alpha\sigma}\phi_{[\mu\nu][\rho\sigma]} .
    \end{equation}
    \item $(2,1)  \oplus (1,2)$: Due to symmetry considerations, we must write the Lorentz structure of the resultant operator as 
    \begin{equation}
    F^{\mu\rho}F^{\nu}_{\rho}F^{\alpha\beta}\phi_{\{ \mu\nu \}[\alpha\beta]} .
    \end{equation}
    \item $(3,0) \oplus (0,3)$: We are again left with two possible Lorentz structures,
    \begin{equation}
        F^{\mu\nu}F^{\rho\sigma}F^{\alpha \beta}\phi_{[\mu\nu][\rho\sigma][\alpha\beta]} \quad \text{and} \quad F^{\mu\nu}F^{\rho\sigma}F^{\alpha \beta}\phi_{[\mu\rho][\nu\sigma][\alpha\beta]}.
    \end{equation}
\end{itemize}
 
\subsubsection{Tri-gluon operators: Color Selection Rules}

We now explore operators that include three gluon field strength tensors. We have decided to present our results using a mix of conventions. Our tables throughout this paper show operators and Clebsch-Gordan coefficients using fundamental SU(3)$_\text{c}$ indices. This is useful for understanding the SU(3)$_\text{c}$ representations. 
However, below we also present arguments about spin-color selection rules using gluon indices. We believe the presentation using gluon indices makes the vanishing of some operators due to color/spin symmetry considerations more transparent. We have checked some of the below results about vanishing coefficients in 
\textsc{FeynRules}~\cite{Christensen:2008py, Alloul:2013bka} and \textsc{MadGraph5\texttt{\textunderscore}aMC@NLO} (\textsc{MG5\texttt{\textunderscore}aMC})~\cite{Alwall:2014hca, Frederix:2018nkq}.  

We determine allowed  LEX SU(3)$_\text{c}$ representations in operators of type $GGG\phi$ by looking to the SU(3)$_\text{c}$ product rule
\begin{align} \label{eq:63}
\begin{split}
8\otimes 8 \otimes 8 &= (\textcolor{red}{27}\oplus \textcolor{blue}{\overline{10}} \oplus \textcolor{orange}{10} \oplus \textcolor{Fuchsia}{8} \oplus \textcolor{Thistle}{8} \oplus \textcolor{Green}{1}) \otimes 8 \\
&= \textcolor{red}{64 \oplus \overline{35} \oplus 35 \oplus 27 \oplus \overline{10} \oplus 27 \oplus 10 \oplus 8} \oplus \textcolor{blue}{\overline{35} \oplus 27 \oplus \overline{10} \oplus 8} \oplus \textcolor{orange}{35 \oplus 27 \oplus 10 \oplus 8}
\\
& \quad \ \ \ \ \ \  \oplus \textcolor{Fuchsia}{27 \oplus \overline{10} \oplus 10 \oplus 8 \oplus 8 \oplus 1} \oplus \textcolor{Thistle}{27 \oplus \overline{10} \oplus 10 \oplus 8 \oplus 8 \oplus 1} \oplus \textcolor{green}{8}.
\end{split}
\end{align}
We have computed the tensor product rules using the method of Coleman~\cite{Coleman:1965afp}. From this, we see that we could potentially reach LEX states with color charges up to a 64-plet---in addition to 35-plets, 27-plets, decouplets, octets, and singlets. 

It is necessary to examine the symmetries of the $8\otimes 8\otimes 8$ tensor product. We can consider, briefly, the tenor product of  three gluons as a tensor with three gluon (adjoint) indices. A rank three tensor can be decomposed as the sum of a totally antisymmetric rank 3 tensor, a totally symmetric rank 3 tensor, and two pieces  of mixed symmetry---one of which is symmetric under interchange of two indices and the other anti-symmetric~\cite{Itin_2021}:
\begin{align}
\begin{split}
T^{abc}=S^{abc}+A^{abc}+N_1^{abc}+N_2^{abc} \\
 N_1^{ijk}=\frac{1}{3}(N^{ijk}+N^{jki}-N^{kij}-N^{kji})\\
 N_2^{ijk}=\frac{1}{3}(N^{ijk}-N^{jki}+N^{kij}-N^{kji}) .
 \end{split}
\end{align}
We can thus write the product of $8\otimes 8\otimes 8$ as an object with three gluon indices and we may place each term in the triple gluon tensor product in the totally symmetric, totally antisymmetric or mixed symmetry parts of the three index object.

In the totally antisymmetric 3 index object we have 56  degrees of freedom ($ \binom{8}{3} =8\times7\times6/6 =56$). These 56 degrees of freedom contain the color representations $A\supset 1, 8, 10, \overline{10},27$. A totally symmetric rank $k$ tensor has $\binom{n+k-1}{k}$ degrees of freedom. This is gives us 120 degrees of freedom ($\binom{10}{3} = 10\times9\times8/6=120$). The totally symmetric part of the product must contain the highest representation, in addition to the singlet. Thus, we can see that the totally symmetric part of the product contains the color representations $ S \supset 64,  27, 10, \overline{10},8 , 1$. 

The remaining pieces of the product must go into mixed symmetry pieces. However, by imagining the Young tableaux, we can convince ourselves that the two parts of mixed symmetry $N_1$ and $N_2$ should equally divide the remaining states in the product. That is $\{N_1, N_2 \}\supset 35,  \overline{35},  27,  27, \overline{10},10,8,8,8$. They have 168 degrees of freedom each. We can see that the only representations given in Eq.~\ref{eq:63} that are not contained in the mixed symmetry products are the 1 and 64.  We can then add the components of $A$, $S$, $N_1$, and $N_2$ to get 512 degrees of freedom, which is correct for a  rank three tensor whose indices can take 8 values.

In studying the phenomenology of potential LEX states, we find some interesting color/spin selection rules from symmetry considerations. From these, we can extract some information about the color flow of these operators.  

\subsubsection{Spin 0 LEX state}
We will briefly dissect the three-gluon operators that involve a spin 0 LEX state.  For scalar LEX states, all Lorentz indices must be contracted between gluon field strength tensors, as discussed in Section~\ref{s3.8.1}: $F_{\mu\nu}F^{\nu}_{~\rho} F^{\rho \mu}\phi$. We begin by showing that the three-gluon operators are completely anti-symmetric under interchange of  gluons in the case of a singlet scalar LEX state. For ease of notation in this section, we choose lower case Roman letters at the beginning of the alphabet to be adjoint SU(3)$_\text{c}$ indices. We begin with the structure
\begin{align}
G_{\mu\nu}^a G^{ b  \nu}_{\rho} G^{c \rho \mu} ,
\end{align}
where $a, b, c$ are the gluon indices. Interchanging gluons $a \leftrightarrow b$, we obtain
\begin{align}
\begin{split}
G_{\mu\nu}^b G^{ a  \nu}_{\rho} G^{c \rho \mu} &= G_{\mu}^{b \nu} G^{ a }_{\rho\nu} G^{c \rho \mu}\\
&= G_{\rho}^{b \nu} G^{ a }_{\mu\nu} G^{c \mu \rho} = - G_{\rho}^{b \nu} G^{ a }_{\mu\nu} G^{c \rho \mu} = - G_{\mu\nu}^a G^{ b  \nu}_{\rho} G^{c \rho \mu},
\end{split}
\end{align}
where between the first and second lines we interchange Lorentz indices $\mu \leftrightarrow \rho$. We could obtain similar statements for the interchange of any other pair of gluons. 

\textbf{Color Singlets:} Due to the anti-symmetry above, for the color singlet LEX scalar state with quantum numbers $(1,1,0)$, we see that the adjoint gluon color indices must contract through the totally antisymmetric $f_{abc}$.

\textbf{ Color octet: } In the case of LEX color octet, we have several options for color contractions of the operator. The possible color contractions between pairs of octets within the operator are
 \begin{align} \label{eq:71}
 \begin{split}
 ( G_{\mu\nu}^a G^{ b  \nu}_{\rho})_1 (G^{c \rho \mu} &\phi^d)_1 , \qquad 
( G_{\mu\nu}^a G^{ b  \nu}_{\rho})_8 (G^{c \rho \mu} \phi^d)_8 , \qquad
( G_{\mu\nu}^a G^{ b  \nu}_{\rho})_{10} (G^{c \rho \mu} \phi^d)_{\overline{10}} , \\
&( G_{\mu\nu}^a G^{ b  \nu}_{\rho})_{\overline{10}} (G^{c \rho \mu} \phi^d)_{10} , \qquad
( G_{\mu\nu}^a G^{ b  \nu}_{\rho})_{\overline{27}} (G^{c \rho \mu} \phi^d)_{27} .
 \end{split}
\end{align}
These color contractions are each orthogonal to one other and so in principle can involve distinct operator coefficients. 

From this, we might then attempt to pull some observable information out of the existence of color contractions.  Color projection techniques have been under ongoing study for multi-gluon processes (for example~\cite{Cougoulic:2025wpl}). We will also attempt to make use of the detailed study of color structure for our exotics to see if we can extract any interesting information. We will now take advantage of symmetry to see that only a few of our possible color projections in Eq.~\ref{eq:71} will survive. 

The  product $8\otimes8$ is an object with two gluon indices. As discussed in Section~\ref{s3.1.2}, this may be decomposed into symmetric and anti-symmetric pieces. 
The symmetric $1\oplus8_s \oplus 27$ objects are symmetric under interchange of two gluon indices. The color structure where two gluons are contracted into a singlet and the LEX state and remaining gluon are contracted into a second singlet can be written as 
\begin{align}
G_{\mu\nu}^a G^{ b  \nu}_{\rho} G^{c \rho \mu} \phi^d  \delta_{ab}\delta_{cd} .
\end{align}
However, our operators must be totally antisymmetric under any two gluon exchanges. These operators must then equal zero due to color projections. As such, the symmetric two gluon contractions vanish. As the color contraction of two gluons into a 27 is also symmetric, only three of the operators listed in Eq.~\ref{eq:71} remain. 
 

When two gluons are contracted into an octet, this can occur through a symmetric or antisymmetric contraction. Similarly, the remaining gluon and octet can be contracted either symmetrically or anti-symmetrically. However, we see again that due to the antisymmetric nature of the operator under the interchange of two gluons, the first pair of gluons must be contracted into an $8_a$ with the antisymmetric $f^{abe}$. Contracting the remaining gluon and LEX state could, at first glance, be symmetric or antisymmetric. However, checking the value of the product $f^{abe}f^{cde}$~\cite{Haber:2019sgz}, we see that this yields zero when contracted with a totally antisymmetric object.  The only non-zero color contraction is then
\begin{align}
G_{\mu\nu}^a G^{ b  \nu}_{\rho} G^{c \rho \mu} \phi^d  f_{eab}d^e_{cd} .
\end{align}

We may contract two gluons into a 10 or $\overline{10}$ representation with an appropriate Clebsch-Gordan coefficient. Diving into the literature, we find that one can project out the antisymmetric 10 and $\overline{10}$ coefficients of a two octet SU(3)$_\text{c}$ tensor product. The $10+\overline{10}$ can separated by subtracting the  the product into its anti-symmetric component and from this subtracting the antisymmetric octet
\begin{equation}
[TT]_{10+\overline{10}}^{ab} =T_1^a T_2^b -T_1^b T_2^a -f^{abe}f^{cde}T_1^cT_2^d .
\end{equation}
We may act on this with the projectors
\begin{equation}
\frac{1}{2}[1 + if^{acg}f^{bch}] [TT]_{10+\overline{10}}~~~~~~ \frac{1}{2}[1 - if^{gia}f^{hib}] [TT]_{10+\overline{10}}^{ab}  
\end{equation}
to extract the $10$ and $\overline{10}$ color contractions~\cite{Dashen:1994qi}. \\

\textbf{LEX states of higher color} 


 \textbf{The 27:} The highest dimensional representation that is in the totally anti-symmetric product of $8\otimes8\otimes 8$ is the 27 of SU(3)$_\text{c}$. Therefore, we can couple a Lorentz scalar LEX 27-plet to three gluon field strength tensors. 

 \textbf{The 35 pseudoscalar:}
The CP even scalar 35-plet cannot couple to three gluon field strength tensors due to symmetry. To couple the 35-plet LEX state, we must add an epsilon tensor to our operator.
The operator then becomes
\begin{equation}
G^a_{\mu\nu}G^{b\nu}_{\rho}\epsilon^{\mu\rho\alpha\beta}G^c_{\alpha\beta}\widetilde{\phi}
\end{equation}
The operator is no longer totally antisymmetric under gluon interchange, and as such a 35 of SU(3)$_\text{c}$ is accessible if it is a CP-odd pseudo-scalar.

\textbf{Is a color 64 accessible?}
Finally, we see that the highest color representation in the tensor product of three octets is the 64. In fundamental indices, this object has three upper and three lower indices, $\phi^{ABC}_{DEF}$. The 64 of SU(3) cannot couple to three gluons if that 64 is a spin 0 Lorentz scalar. This is because the 64 is in the totally symmetric part of the  $8\otimes8\otimes 8$ product. It also cannot couple as a pseudo-scalar since it is in the totally symmetric part of the gluon triple product.  As such, we find that LEX states in 64 dimensional representation must be at least in the spin 1 Lorentz representation in order to couple to three SU(3)$_\text{c}$ field strength tensors. It is a curious coincidence of operator symmetry that a spin 0 64-plet is forbidden by color contraction.

We present operators of type $GGG\phi$, where $\phi$ is the lowest spin available, in Table~\ref{tab:8}. For all color representations except the 64-plet, the LEX state in Table~\ref{tab:8} is a scalar. 


\begin{table}[h!]
\begin{center}
\begin{tabular}{|c|c|c|c|}
\hline 
CG Coefficients & Operators & (SU(3)$_{\text{c}}$, SU(2)$_{\text{L}}$, U(1)$_Y$) & Portal \tabularnewline
\hline 
\hline  
${C_{64,8,8,8}}^{BDFJKL}_{ACEGHI}$ & $G^{\mu\nu A}_B G^C_{\mu\nu D} G^{\sigma\rho E}_{\ F} \phi^{GHI}_{JKL \sigma\rho}$ & ($64,1,0$) & $\diamond$ \tabularnewline 
\hline  
${C_{35,8,8,8}}^{BDFK}_{ACEGHIJ}$ & $G^{\mu\nu A}_B \tilde{G}^C_{\mu\sigma D} G^{\sigma E}_{\nu F} \tilde{\phi}^{GHIJ}_{K}$ & ($35,1,0$) & $\diamond$  \tabularnewline 
\hline  
${C_{27,8,8,8}}^{BDFIJ}_{ACEGH}$ & $G^{\mu\nu A}_B G^C_{\mu\sigma D} G^{\sigma E}_{\nu F} \phi^{GH}_{IJ}$ & ($27,1,0$)  & $\diamond$ \tabularnewline 
\hline  
${C_{10,8,8,8}}^{BDF}_{ACEGHI}$ & $G^{\mu\nu A}_B G^C_{\mu\sigma D} G^{\sigma E}_{\nu F} \phi^{GHI}_{}$ & ($10,1,0$)  & $\diamond$ \tabularnewline 
\hline  
${C_{8,8,8,8}}^{BDFH}_{ACEG}$ & $G^{\mu\nu A}_B G^C_{\mu\sigma D} G^{\sigma E}_{\nu F} \phi^{G}_{H}$ & ($8,1,0$)  & $\diamond$ \tabularnewline 
\hline  
${C_{8,8,8}}^{BDF}_{ACE}$ & $G^{\mu\nu A}_B G^C_{\mu\sigma D} G^{\sigma E}_{\nu F} \phi^{}_{}$ & ($1,1,0$)  & $\diamond$ 
\tabularnewline 
\hline
\end{tabular}
\end{center}
\caption{\label{tab:8}Dimension 7 operators of type $GGG\phi$ for lowest Lorentz representation LEX states.}
\end{table}

\subsubsection{$(1,0)\oplus(0,1)$ LEX state}
We may look at color-spin selection rules for this spin 1 LEX state. We can once again check the symmetry properties of allowed operators under gluon interchange. First,
\begin{equation}
G^a_{\mu\nu}G^{b\mu\nu}G^c_{\alpha\beta}\phi^{\alpha\beta}_{abc}
\end{equation}
is symmetric under $\alpha\leftrightarrow \beta$. This means we can couple spin 1 LEX states in the symmetric ($S$), and one mixed symmetry ($N_1$) portions of the triple gluon tensor product. The LEX 64-plet is thus accessible, as is the 35 and other representations. Forbidden representations are in the totally anti-symmetric part of the triple gluon product ($A$) and the other mixed symmetry parts ($N_2$).
The operator
\begin{equation}
G^a_{\mu\nu}G^{b\nu}_{\rho}G^{c\rho}_{\sigma}\phi^{\mu\sigma}_{abc}
\end{equation}
is neither symmetric nor anti-symmetric under gluon interchange.

\subsubsection{$(2,0)\oplus(0,2)$ representation}
We next check the color-spin selection rules for the spin 2 object. We start with the operator
\begin{equation}
G^a_{\mu\nu}G^{b\nu}_{\rho}G^c_{\alpha\beta}\phi^{\mu\rho\alpha\beta}_{abc} .
\end{equation}
We can then check that this is antisymmetric under interchange $a\leftrightarrow b$: 
\begin{align}
\begin{split}
G^b_{\mu\nu}G^{a\nu}_{\rho}G^c_{\alpha\beta}\phi^{\mu\rho\alpha\beta}_{abc}=G_{\mu}^{b \nu}G^{a}_{\rho\nu}G^c_{\alpha\beta}\phi^{\mu\rho\alpha\beta}_{abc}  \qquad\mu\leftrightarrow\rho \\
G^a_{\mu\nu}G^{b\nu}_{\rho}G^c_{\alpha\beta}\phi^{\rho\mu\alpha\beta}=-G^a_{\mu\nu}G^{b\nu}_{\rho}G^c_{\alpha\beta}\phi^{\mu\rho\alpha\beta}_{abc} .
\end{split}
\end{align}
We see that $\phi^{\mu\rho\alpha\beta}$ is antisymmetric under the interchange $\mu\leftrightarrow \rho, \alpha\leftrightarrow\beta$. The operator is neither symmetric nor antisymmetric under other gluon index interchanges.

Similar arguments can be made for the $(2,1)\oplus(1,2)$ operators. There, the LEX state is a four index object $\phi^{\mu\rho\alpha\beta}$ which is symmetric in ${\mu\rho}$ and antisymmetric in ${\alpha\beta}$. Thus, we can see that the LEX states $(2,0)\oplus(0,2)$ and $(2,1)\oplus(1,2)$ must be in opposite mixed symmetry representations of the triple gluon tensor product. Both of these operators forbid both color singlet LEX states and color 64 states.

\subsubsection{$(3,0)+(0,3)$}
For LEX states in the spin 3 representation $(3,0)\oplus(0,3)$, we have the three gluon operator
\begin{equation}
G^a_{\mu\nu}G^b_{\rho\sigma}G^c_{\alpha\beta}\phi^{\mu\nu\rho\sigma\alpha\beta}_{abc} .
\end{equation}
We can show this operator is symmetric under permutation of gluons. For example, under $a\leftrightarrow b$,
\begin{equation}
G^b_{\mu\nu}G^a_{\rho\sigma}G^c_{\alpha\beta}\phi^{\mu\nu\rho\sigma\alpha\beta}_{abc} =
G^a_{\mu\nu}G^b_{\rho\sigma}G^c_{\alpha\beta}\phi^{\rho\sigma\mu\nu\alpha\beta}_{abc}=G^a_{\mu\nu}G^b_{\rho\sigma}G^c_{\alpha\beta}\phi^{\mu\nu\rho\sigma\alpha\beta}_{abc}
\end{equation}
where in the first equality we interchanged indices  $\mu\nu \leftrightarrow\rho\sigma$.
As such, the SU(3)$_\text{c}$ representations of $\phi$ that appear in the operator must be those in the totally symmetric triple gluon product ($S\supset64,27,\overline{10},10,8,1$), but not the 35. 

\subsubsection{GGF operators} 
We note in Table~\ref{tab:9} the presence of many operators that contain only one or two SU(3)$_\text{c}$ field strength tensors (in conjunction with two or one other SM field strength tensors, respectively).  These operators have their own color-spin selection rules. In this section, we will discuss the operators with two gluon field strength tensors.

We can see some color-spin selection rules for operators of the type $GGF\phi$. For example, for a spin 0 LEX state we see that the operator $G^a_{\mu\nu}G^{b\nu}_{\rho}W^{\mu\rho}\phi$ is antisymmetric under exchange of gluons. Therefore, the field $\phi$ must be in the antisymmetric color representations of the $8\otimes8$ product, that is the $10$ or $\overline{10}$. In contrast,  a spin 1 LEX state can interact through the operator $G^a_{\mu\nu}G^{b\mu}_{\nu}W^{\rho\sigma}\phi_{\rho\sigma}$. This operator is symmetric under the interchange of two gluons, and thus the LEX state must be in a representation in the symmetric product of $8\otimes8$ (the 27, $8_s$, or 1). The spin 3 LEX state with six indices also symmetric under gluon interchange. 

These operators allow the production of high color-charged LEX states in association with electroweak gauge bosons.  For example, one possible process is the production of a color 27 through $pp\rightarrow W \phi$. The 27-plet may also be produced in association with a hard photon through $pp\rightarrow \gamma \phi$.  These operators also allow for the three body decays of LEX states into two gluons and an electroweak gauge boson through  $\phi\rightarrow g g+V$.

\subsubsection{GFF operators} 
Finally, we consider operators where a LEX state couples to one SU(3)$_\text{c}$ field strength tensor and two other SM field strength tensors. LEX states in these operators are color octets. These perators have the phenomenological feature of accessing LEX color octets in very high SU(2)$_\text{L}$ representations. The operators with two $W$ bosons allow for the production of LEX states that are up to a quintuplet of SU(2)$_\text{L}$, with electric charges $(2,1,0,-1,-2)$. Some possible production processes include $qq\rightarrow\phi W g$ and $gg\rightarrow WW\phi$. LEX states have three body decays to a gluon and two electroweak gauge bosons, $\phi\rightarrow g + V V$. Doubly charged octets would  decay to two $W$ bosons, potentially creating an intriguing di-lepton signature. 
\begin{table}[h!]
\begin{center}
\begin{tabular}{|c|c|c|c|}
\hline 
CG Coefficients & Operators & (SU(3)$_{\text{c}}$, SU(2)$_{\text{L}}$, U(1)$_Y$) & Portal \tabularnewline
\hline 
\hline 
${C_{27,8,8}}^{BDGH}_{ACEF}$  & $G^{\mu\nu A}_B G_{\mu\sigma D}^C W^\sigma_\nu \phi^{EF}_{GH}$ & ($27,3,0$)  & $\diamond$ \tabularnewline 
\hline  
${C_{10,8,8}}^{BD}_{ACEFG}$ & $G^{\mu\nu A}_B \widetilde{G}_{\mu\sigma D}^C W^\sigma_\nu \widetilde{\phi}^{EFG}_{}$ & ($10,3,0$)  & $\diamond$ \tabularnewline 
\hline  
${C_{\overline{10},8,8}}^{BDEFG}_{AC}$ & $G^{\mu\nu A}_B \widetilde{G}_{\mu\sigma D}^C W^\sigma_\nu \widetilde{\phi}_{EFG}^{}$ & ($\overline{10},3,0$)  & $\diamond$ \tabularnewline 
\hline  
${C_{8,8,8}}^{BDF}_{ACE}$ & $G^{\mu\nu A}_B G_{\mu\sigma D}^C W^\sigma_\nu \phi_{F}^{E}$ & ($8,3,0$)  & $\diamond$ \tabularnewline 
\hline  
${C_{8,8}}^{BD}_{AC}$ & $G^{\mu\nu A}_B G_{\mu\sigma D}^C W^\sigma_\nu \phi_{}^{}$ & ($1,3,0$)  & $\diamond$ \tabularnewline 
\hline  
${C_{27,8,8}}^{BDGH}_{ACEF}$ & $G^{\mu\nu A}_B G_{\mu\sigma D}^C B^\sigma_\nu \phi^{EF}_{GH}$ & ($27,1,0$)  & $\diamond$ \tabularnewline 
\hline  
${C_{10,8,8}}^{BD}_{ACEFG}$ & $G^{\mu\nu A}_B \widetilde{G}_{\mu\sigma D}^C B^\sigma_\nu \widetilde{\phi}^{EFG}_{}$ & ($10,1,0$)  & $\diamond$ \tabularnewline 
\hline  
${C_{\overline{10},8,8}}^{BDEFG}_{AC}$ & $G^{\mu\nu A}_B \widetilde{G}_{\mu\sigma D}^C B^\sigma_\nu \widetilde{\phi}_{EFG}^{}$ & ($\overline{10},1,0$)  & $\diamond$ \tabularnewline 
\hline  
${C_{8,8,8}}^{BDF}_{ACE}$ & $G^{\mu\nu A}_B G_{\mu\sigma D}^C B^\sigma_\nu \phi_{F}^{E}$ & ($8,1,0$)  & $\diamond$ \tabularnewline 
\hline  
${C_{8,8}}^{BD}_{AC}$ & $G^{\mu\nu A}_B G_{\mu\sigma D}^C B^\sigma_\nu \phi_{}^{}$ & ($1,1,0$)  & $\diamond$ \tabularnewline 
\hline  
${C_{8,8}}^{BD}_{AC}$ & $G^{\mu\nu A}_B W_{\mu\sigma} W^\sigma_\nu \phi^C_D$ & ($8,5,0$)  & $\diamond$ \tabularnewline 
\hline
${C_{8,8}}^{BD}_{AC}$ & $G^{\mu\nu A}_B W_{\mu\sigma} W^\sigma_\nu \phi^C_D$ & ($8,3,0$) & $\diamond$  \tabularnewline 
\hline
${C_{8,8}}^{BD}_{AC}$ & $G^{\mu\nu A}_B W_{\mu\sigma} W^\sigma_\nu \phi^C_D$ & ($8,1,0$)  & $\diamond$ \tabularnewline 
\hline
${C_{8,8}}^{BD}_{AC}$ & $G^{\mu\nu A}_B W_{\mu\sigma} B^\sigma_\nu \phi^C_D$ & ($8,3,0$)  & $\diamond$ \tabularnewline 
\hline
${C_{8,8}}^{BD}_{AC}$ & $G^{\mu\nu A}_B B_{\mu\sigma} B^\sigma_\nu \phi^C_D$ & ($8,1,0$)  & $\diamond$ \tabularnewline 
\hline
\end{tabular}
\end{center}
\caption{\label{tab:9}Dimension 7 operators of type $GFF\phi$, where at least one of the field strength tensors is not the SU(3)$_\text{c}$ field strength for lowest dimension Lorentz representation of LEX state}
\end{table}

\subsection{Note on more Dimension 7 Bosonic Operators}
Other dimension seven operator types in the $g$-$g$ portal are $GFHH\phi$, $GFDD\phi$, $GDD\phi HH$, $GDDDD\phi$, $Gff\phi H$, $GffD\phi$, $ffDD\phi H$, and $ffDDD\phi$. The last four of these operator types would also fall into the $g$-$q$ portal. Additionally, the $g$-$q$ portal would contain operators of type $ffD\phi HH$. We note that some of the operator types listed in this paragraph might be redundant with each other or already-discussed operator types, but we do not work out the entire dimension seven catalog here.

\section{Spin 1/2 LEX States}\label{s4}
We continue by enumerating all operators in the gluon-gluon and gluon-quark portals which involve a fermionic LEX field. 

\subsection{Operators of form $G\psi f$ and $G\psi f H$}
We begin this section by examining dimension five operators which include an SU(3)$_\text{c}$ field, a SM fermion, and a LEX fermion. At dimension six, operators  may  include an additional Higgs field. Due to the presence of the strong field strength tensor, all operators in this section fall into the $g$-$g$ portal. If the SM fermion in the operator is a quark, the operators also fall under the $g$-$q$ portal.  We list  operators of this type in Table~\ref{tab:10} and Table~\ref{tab:11} for spin 1/2 LEX states.  In these tables, we continue to mark the portal(s) where each operator can be found, following the convention established in Section~\ref{s2}.
\begin{table}[h!]
\begin{center}
\begin{tabular}{|c|c|c|c|}
\hline 
CG Coefficients & Operators & (SU(3)$_{\text{c}}$, SU(2)$_{\text{L}}$, U(1)$_Y$) & Portal \tabularnewline
\hline 
\hline 
${C_{15,8,\overline{3}}}^{BCF}_{ADE}$ & $G^{\mu\nu A}_B \overline{Q_L}^i_C \sigma_{\mu\nu} \psi^{DE}_{iF}$ & ($15,2,1/6$)  & $\Cap$ \tabularnewline 
\hline 
${C_{8,\overline{6},\overline{3}}}^{BCDE}_{A}$ & $G^{\mu\nu A}_B \overline{Q_L}^i_C \sigma_{\mu\nu} \psi_{i DE}$ & ($\overline{6},2,1/6$)   & $\Cap$ \tabularnewline 
\hline 
${C_{8,\overline{3},3}}^{BC}_{AD}$ & $G^{\mu\nu A}_B \overline{Q_L}^i_C \sigma_{\mu\nu} \psi_i^D$ & ($3,2,1/6$)  & $\Cap$  \tabularnewline 
\hline 
${C_{15,8,\overline{3}}}^{BCF}_{ADE}$ & $G^{\mu\nu A}_B \overline{u}_C \sigma_{\mu\nu} \psi^{DE}_F$ & ($15,1,2/3$)   & $\Cap$ \tabularnewline 
\hline  
${C_{8,\overline{6},\overline{3}}}^{BCDE}_{A}$ & $G^{\mu\nu A}_B \overline{u}_C \sigma_{\mu\nu} \psi_{DE}$ & ($\overline{6},1,2/3$)  & $\Cap$  \tabularnewline 
\hline  
${C_{8,\overline{3},3}}^{BC}_{AD}$ & $G^{\mu\nu A}_B \overline{u}_C \sigma_{\mu\nu} \psi^D$ & ($3,1,2/3$)   & $\Cap$ \tabularnewline 
\hline  
${C_{15,8,\overline{3}}}^{BCF}_{ADE}$ & $G^{\mu\nu A}_B \overline{d}_C \sigma_{\mu\nu} \psi^{DE}_F$ & ($15,1,-1/3$)   & $\Cap$ \tabularnewline 
\hline
${C_{8,\overline{6},\overline{3}}}^{BCDE}_{A}$ & $G^{\mu\nu A}_B \overline{d}_C \sigma_{\mu\nu} \psi_{DE}$ & ($\overline{6},1,-1/3$)  & $\Cap$  \tabularnewline 
\hline
${C_{8,\overline{3},3}}^{BC}_{AD}$ & $G^{\mu\nu A}_B \overline{d}_C \sigma_{\mu\nu} \psi^D$ & ($3,1,-1/3$)   & $\Cap$ \tabularnewline 
\hline
${C_{8,8}}^{BD}_{AC}$ & $G^{\mu\nu A}_B \overline{L}^i \sigma_{\mu\nu} \psi^C_{iD}$ & ($8,2,-1/2$)  & $\diamond$ \tabularnewline 
\hline
${C_{8,8}}^{BD}_{AC}$ & $G^{\mu\nu A}_B \overline{\ell} \sigma_{\mu\nu} \psi^C_D$ & ($8,1,-1$)  & $\diamond$ \tabularnewline 
\hline
\end{tabular}
\end{center}
\caption{\label{tab:10}Dimension 5 operators of type $G\psi f$ for spin 1/2 LEX states}
\end{table}

\begin{table}[h!]
\begin{center}
\begin{tabular}{|c|c|c|}
\hline 
Operators & (SU(3)$_{\text{c}}$, SU(2)$_{\text{L}}$, U(1)$_Y$) & Portal \tabularnewline
\hline 
\hline 
$G^{\mu\nu} \overline{Q_L}^i \sigma_{\mu\nu} \psi_i^j H_j$ & ($15,3,-1/3$), ($\overline{6},3,-1/3$), ($3,3,-1/3$)   & $\Cap$ \tabularnewline 
\hline  
$G^{\mu\nu} \overline{Q_L}^i \sigma_{\mu\nu} \psi H_i$ & ($15,1,-1/3$), ($\overline{6},1,-1/3$), ($3,1,-1/3$)  & $\Cap$ \tabularnewline 
\hline  
$G^{\mu\nu} \overline{u} \sigma_{\mu\nu} \psi^i H_i$ & ($15,2,1/6$), ($\overline{6},2,1/6$), ($3,2,1/6$)   & $\Cap$  \tabularnewline 
\hline  
$G^{\mu\nu} \overline{d} \sigma_{\mu\nu} \psi^i H_i$ & ($15,2,-5/6$), ($\overline{6},2,-5/6$), ($3,2,-5/6$)   & $\Cap$ \tabularnewline 
\hline
$G^{\mu\nu} \overline{L}^i \sigma_{\mu\nu} \psi^j_i H_j$ & ($8,3,-1$)  & $\diamond$ \tabularnewline 
\hline
$G^{\mu\nu} \overline{L}^i \sigma_{\mu\nu} \psi H_i$ & ($8,1,-1$)  & $\diamond$ \tabularnewline 
\hline
$G^{\mu\nu} \overline{\ell} \sigma_{\mu\nu} \psi^i H_i$ & ($8,2,-3/2$)  & $\diamond$ \tabularnewline 
\hline
$G^{\mu\nu} \overline{Q_L}^i \sigma_{\mu\nu} \psi_{ij} H^{\dagger j}$ & ($15,3,2/3$), ($\overline{6},3,2/3$), ($3,3,2/3$)  & $\Cap$ \tabularnewline 
\hline 
$G^{\mu\nu} \overline{Q_L}^i \sigma_{\mu\nu} \psi H^\dagger_i$ & ($15,1,2/3$), ($\overline{6},1,2/3$), ($3,1,2/3$) & $\Cap$  \tabularnewline 
\hline 
$G^{\mu\nu} \overline{u} \sigma_{\mu\nu} \psi_i H^{\dagger i}$ & ($15,2,7/6$), ($\overline{6},2,7/6$), ($3,2,7/6$) & $\Cap$  \tabularnewline 
\hline
$G^{\mu\nu} \overline{d} \sigma_{\mu\nu} \psi_i H^{\dagger i}$ & ($15,2,1/6$), ($\overline{6},2,1/6$), ($3,2,1/6$)  & $\Cap$ \tabularnewline 
\hline
$G^{\mu\nu} \overline{L}^i \sigma_{\mu\nu} \psi_{ij} H^{\dagger j}$ & ($8,3,0$) & $\diamond$  \tabularnewline 
\hline
$G^{\mu\nu} \overline{L}^i \sigma_{\mu\nu} \psi H^\dagger_i$ & ($8,1,0$) & $\diamond$  \tabularnewline 
\hline
$G^{\mu\nu} \overline{\ell} \sigma_{\mu\nu} \psi_i H^{\dagger i}$ & ($8,2,-1/2$) & $\diamond$  \tabularnewline 
\hline
\end{tabular}
\end{center}
\caption{\label{tab:11}Dimension 6 operators of type $G\psi f H$ for spin 1/2 LEX states. SU(3)$_\text{c}$ indices are not written out.}
\end{table}

In the operators of Table~\ref{tab:10} and Table~\ref{tab:11} that contain no SM quarks, the LEX state are color octets. In the operators with a SM quark present, we need the SU(3)$_\text{c}$ tensor product given in Eq.~\ref{eq:3}. The LEX state can be a color triplet, sextet, or 15-plet. 

The SU(2)$_\text{L}$ charge of the LEX state is dependent on the presence of a Higgs boson and whether the SM fermion is left- or right-handed. In the case of no Higgs present, the LEX state is an electroweak singlet if there is a right-handed SM fermion, and an electroweak doublet if there is a left-handed SM fermion. In operators that contain a Higgs field and a right-handed SM fermion, the LEX state is similarly an electroweak doublet. Finally, in operators that contain both a Higgs state and a left-handed fermion, the LEX state can be an SU(2)$_\text{L}$ triplet or singlet.
Among the most interesting phenomenological signatures resultant from Table~\ref{tab:11} is the production of SU(2)$_\text{L}$ triplet or doublet LEX states from quark-gluon fusion, and the three body decays of these LEX states.

\subsubsection{Higher spin LEX states in $Gf\psi$ operators}
We can determine allowed Lorentz representations of higher spin LEX states  with the operator tensor product 
\begin{align}
\begin{split}
[(1,0)\oplus(0,1)]&\otimes [(1/2,0)\oplus(0,1/2)] \\
&= [(3/2,0)\oplus(0,3/2)]\oplus[(1/2,0)\oplus(0,1/2)] \oplus [(1,1/2)\oplus(1/2,1)] .
\end{split}
\end{align}
We find that in addition to a standard spin 1/2 fermion, the LEX state may be a spin 3/2 object taking one of two representations. The $[(1,1/2)\oplus (1/2,1)]$ object is the Rarita-Schwinger field. Thus there is an operator with Lorentz structure
\begin{equation}
G^{\mu\nu}\overline{f}\gamma_{\mu}\psi_{\nu} .
\end{equation}
The $[(3/2,0)\oplus(0,3/2)]$ representation can be found in operators of the form 
\begin{equation}
G^{\mu\nu}\overline{f}\psi_{[\mu\nu]} .
\end{equation}

\subsection{Operators of Form $D\psi f$, $D\psi f H$, and $D\psi f HH$} \label{s3.2}
The next operator types that we consider contain one covariant derivative, a SM fermion and a LEX fermion. These operators may contain zero, one, or two Higgs fields. These operators will contain a single gamma matrix, in order to ``soak up" the Lorentz index of the covariant derivative. Operators of this type contain only one derivative and no field strength tensors; as such, they fall into the gluon-quark portal but not the gluon-gluon portal. We note that the SM fermion is required to be a quark.

We can start by analyzing the simplest of these operators, without any Higgs fields. In this case, there are two options for derivative placement, which are directly related through integration by parts. This leaves us with only one operator, which we choose as $D_\mu \overline{q} \gamma^\mu \psi$. We note that the inclusion of Higgs fields does not increase the number of operators in the portal, as they are not charged under SU(3)$_\text{c}$. All operators of this type are shown in Table~\ref{tab:12} for spin 1/2 LEX states.
\begin{table}[h!]
\begin{center}
\begin{tabular}{|c|c|c|c|}
\hline 
Dimension & Operators & (SU(3)$_{\text{c}}$, SU(2)$_{\text{L}}$, U(1)$_Y$) & Portal \tabularnewline
\hline 
\hline 
dim 4 & $D_\mu \overline{Q_L}^i \gamma^\mu \psi_i$ & ($3,2,1/6$) & $\star$ \tabularnewline 
\hline  
dim 4 & $D_\mu \overline{u} \gamma^\mu \psi$ & ($3,1,2/3$) & $\star$ \tabularnewline 
\hline  
dim 4 & $D_\mu \overline{d} \gamma^\mu \psi$ & ($3,1,-1/3$) & $\star$ \tabularnewline 
\hline  
dim 5 & $D_\mu \overline{Q_L}^i \gamma^\mu \psi_i^j H_j$ & ($3,3,-1/3$) & $\star$ \tabularnewline 
\hline 
dim 5 & $D_\mu \overline{Q_L}^i \gamma^\mu \psi_{ij} H^{\dagger j}$ & ($3,3,2/3$) & $\star$ \tabularnewline 
\hline 
dim 5 & $D_\mu \overline{Q_L}^i \gamma^\mu \psi H_i$ & ($3,1,-1/3$) & $\star$ \tabularnewline 
\hline 
dim 5 & $D_\mu \overline{Q_L}^i \gamma^\mu \psi H^\dagger_i$ & ($3,1,2/3$) & $\star$ \tabularnewline 
\hline 
dim 5 & $D_\mu \overline{u} \gamma^\mu \psi^i H_i$ & ($3,2,1/6$) & $\star$ \tabularnewline 
\hline  
dim 5 & $D_\mu \overline{u} \gamma^\mu \psi_i H^{\dagger i} $ & ($3,2,7/6$) & $\star$ \tabularnewline 
\hline  
dim 5 & $D_\mu \overline{d} \gamma^\mu \psi^i H_i$ & ($3,2,-5/6$) & $\star$ \tabularnewline 
\hline  
dim 5 & $D_\mu \overline{d} \gamma^\mu \psi_i H^{\dagger i} $ & ($3,2,1/6$) & $\star$ \tabularnewline 
\hline  
dim 6 & $D_\mu \overline{Q_L}^i \gamma^\mu \psi_i^{jk} H_j H_k$ & ($3,4,-5/6$) & $\star$ \tabularnewline 
\hline 
dim 6 & $D_\mu \overline{Q_L}^i \gamma^\mu \psi_{ij}^{k} H^{\dagger j} H_k$ & ($3,4,1/6$) & $\star$ \tabularnewline 
\hline  
dim 6 & $D_\mu \overline{Q_L}^i \gamma^\mu \psi_{ijk}^{} H^{\dagger j} H^{\dagger k} $ & ($3,4,7/6$) & $\star$ \tabularnewline 
\hline  
dim 6 & $D_\mu \overline{Q_L}^i \gamma^\mu \psi^j H_i H_j$ & ($3,2,-5/6$) & $\star$ \tabularnewline 
\hline  
dim 6 & $D_\mu \overline{Q_L}^i \gamma^\mu \psi_j H_i H^{\dagger j}$, \ $D_\mu \overline{Q_L}^i \gamma^\mu \psi_i H_j H^{\dagger j}$ & ($3,2,1/6$) & $\star$ \tabularnewline 
\hline 
dim 6 & $D_\mu \overline{Q_L}^i \gamma^\mu \psi_j H^\dagger_i H^{\dagger j}$ & ($3,2,7/6$) & $\star$ \tabularnewline 
\hline 
dim 6 & $D_\mu \overline{u} \gamma^\mu \psi^{ij} H_i H_j$ & ($3,3,-1/3$) & $\star$ \tabularnewline 
\hline
dim 6 & $D_\mu \overline{u} \gamma^\mu \psi^{i}_j H_i H^{\dagger j} $ & ($3,3,2/3$) & $\star$ \tabularnewline 
\hline
dim 6 & $D_\mu \overline{u} \gamma^\mu \psi_{ij} H^{\dagger i} H^{\dagger j} $ & ($3,3,5/3$) & $\star$ \tabularnewline 
\hline
dim 6 & $D_\mu \overline{u} \gamma^\mu \psi H_i H^{\dagger i} $ & ($3,1,2/3$) & $\star$ \tabularnewline 
\hline
dim 6 & $D_\mu \overline{d} \gamma^\mu \psi^{ij} H_i H_j$ & ($3,3,-4/3$) & $\star$ \tabularnewline 
\hline
dim 6 & $D_\mu \overline{d} \gamma^\mu \psi^{i}_j H_i H^{\dagger j} $ & ($3,3,-1/3$) & $\star$ \tabularnewline 
\hline
dim 6 & $D_\mu \overline{d} \gamma^\mu \psi_{ij} H^{\dagger i} H^{\dagger j} $ & ($3,3,2/3$) & $\star$ \tabularnewline 
\hline
dim 6 & $D_\mu \overline{d} \gamma^\mu \psi H_i H^{\dagger i} $ & ($3,1,-1/3$) & $\star$ \tabularnewline 
\hline
\end{tabular}
\end{center}
\caption{\label{tab:12}
Operators of type $D\psi f$, $D\psi f H$, and $D\psi f HH$. SU(3)$_\text{c}$ indices are not written out. Operator dimension is included.}
\end{table}

Within the operators of Table~\ref{tab:12}, the only SM field charged under SU(3)$_\text{c}$ is a singular anti-quark. As such, all of the LEX fermions are color triplets. Due to this simplicity, we have not chosen to write out the SU(3) indices of the states in Table~\ref{tab:12}. As before, in operators which only contain SM fields that are electroweak singlets, the LEX state is also an electroweak singlet, while operators containing one SM electroweak doublet also have a LEX state that is an electroweak doublet. In the case that two of the SM fields are SU(2) doublets, either with two Higgs fields and a right-handed SM fermion or with one Higgs field and a left-handed SM fermion, the LEX state will be an SU(2)$_\text{L}$ singlet or triplet. Finally, the maximal case is operators which contain three SM fields, all of which are electroweak doublets. In this case, the LEX field will be an SU(2) quadruplet or doublet.

The operators in Table~\ref{tab:12} represent various types of exotic quarks.  Perhaps the most interesting are the weak quadruplets which contain states of higher charge.  

\subsubsection{Higher spin LEX states} Looking at this operator, we can take the Lorentz tensor product
\begin{align}
(1/2,1/2)\otimes[(1/2,0)\oplus(0,1/2)]=[(1,1/2)\oplus(1/2,1)]\oplus[(1/2,0)\oplus(0,1/2)]
\end{align}
to see that the higher spin LEX field associated with this operator is the Rarita Schwinger field with operator $D^{\mu}\overline{f}\psi_{\mu}$.

\subsection{Operators of form $FD \psi f$ }
Next, we examine operators that include one derivative, along with a field strength tensor, LEX fermion, and SM fermion. In order for operators of this type to fall into the gluon-gluon portal, there must be an SU(3)$_\text{c}$ field strength tensor. However, within this portal there may be any choice of SM fermion. Meanwhile, for operators to fall into the gluon-quark portal, we may include any SM field strength tensor, but the SM fermion must be a quark.

Within the gluon-gluon portal, the derivative may act on any of the fields, giving us three operators that are related through one integration by parts relation:
\begin{equation}
    \overline{f}\gamma^\mu \overleftarrow{D^\nu} \psi G_{\mu \nu} = 
    \overline{Q}\gamma^\mu \overrightarrow{D^\nu} \psi G_{\mu \nu} +
    \overline{Q}\gamma^\mu \psi D^\nu G_{\mu \nu} .
\end{equation}
We are thus left with two independent operators of this type in the $g$-$g$ portal. The $g$-$q$ portal, however, is more complicated. In the case that there is an SU(3)$_\text{c}$ field strength tensor, the above reasoning still holds. However, in the case that a different field strength tensor is present, the initial number of operators in the portal is decreased by one. In this case, there will only be one remaining independent operator within the $g$-$q$ portal. Operators of this type, in either of the two portals, are listed in Table~\ref{tab:13} for spin 1/2 LEX states.

\begin{table}[h!]
\begin{center}
\begin{tabular}{|c|c|c|c|}
\hline 
CG Coefficients & Operators & (SU(3)$_{\text{c}}$, SU(2)$_{\text{L}}$, U(1)$_Y$) & Portal \tabularnewline
\hline 
\hline 
\multirow{2}{*}{${C_{15,8,\overline{3}}}^{BCF}_{ADE}$} & $G^{\mu\nu A}_B \overline{Q_L}^i_C \gamma_{\mu} D_\nu \psi^{DE}_{iF},  $ & \multirow{2}{*}{($15,2,1/6$)} & \multirow{2}{*}{$\Cap$} \tabularnewline 
& $G^{\mu\nu A}_B D_\nu \overline{Q_L}^i_C \gamma_{\mu} \psi^{DE}_{iF}$ & & \tabularnewline
\hline  
\multirow{2}{*}{${C_{8,\overline{6},\overline{3}}}^{BCDE}_{A}$} & $G^{\mu\nu A}_B \overline{Q_L}^i_C \gamma_{\mu} D_\nu \psi_{iDE},  $ & \multirow{2}{*}{($\overline{6},2,1/6$)} & \multirow{2}{*}{$\Cap$} \tabularnewline 
& $G^{\mu\nu A}_B D_\nu \overline{Q_L}^i_C \gamma_{\mu} \psi_{iDE}$ & & \tabularnewline
\hline  
\multirow{2}{*}{${C_{8,\overline{3},3}}^{BC}_{AD}$} & $G^{\mu\nu A}_B \overline{Q_L}^i_C \gamma_{\mu} D_\nu \psi_i^D,  $ & \multirow{2}{*}{($3,2,1/6$)} & \multirow{2}{*}{$\Cap$} \tabularnewline 
& $G^{\mu\nu A}_B D_\nu \overline{Q_L}_C^i \gamma_{\mu} \psi_i^D$ & & \tabularnewline
\hline  
${C_{15,8,\overline{3}}}^{BCF}_{ADE}$ & $G^{\mu\nu A}_B \overline{u}_C \gamma_{\mu}D_{\nu} \psi^{DE}_F, \quad G^{\mu\nu A}_B D_{\nu} \overline{u}_C \gamma_{\mu}\psi^{DE}_F$ & ($15,1,2/3$) & $\Cap$ \tabularnewline 
\hline  
${C_{8,\overline{6},\overline{3}}}^{BCDE}_{A}$ & $G^{\mu\nu A}_B \overline{u}_C \gamma_{\mu}D_{\nu} \psi_{DE}, \quad G^{\mu\nu A}_B D_{\nu} \overline{u}_C \gamma_{\mu}\psi_{DE}$ & ($\overline{6},1,2/3$) & $\Cap$ \tabularnewline 
\hline  
${C_{8,\overline{3},3}}^{BC}_{AD}$ & $G^{\mu\nu A}_B \overline{u}_C \gamma_{\mu}D_{\nu} \psi^D, \quad G^{\mu\nu A}_B D_{\nu} \overline{u}_C \gamma_{\mu}\psi^D$ & ($3,1,2/3$) & $\Cap$ \tabularnewline 
\hline  
${C_{15,8,\overline{3}}}^{BCF}_{ADE}$  & $G^{\mu\nu A}_B \overline{d}_C \gamma_{\mu}D_{\nu} \psi^{DE}_F, \quad G^{\mu\nu A}_B D_{\nu} \overline{d}_C \gamma_{\mu} \psi^{DE}_F$ & ($15,1,-1/3$) & $\Cap$ \tabularnewline 
\hline
${C_{8,\overline{6},\overline{3}}}^{BCDE}_{A}$ & $G^{\mu\nu A}_B \overline{d}_C \gamma_{\mu}D_{\nu} \psi_{DE}, \quad G^{\mu\nu A}_B D_{\nu} \overline{d}_C \gamma_{\mu} \psi_{DE}$ & ($\overline{6},1,-1/3$) & $\Cap$ \tabularnewline 
\hline
${C_{8,\overline{3},3}}^{BC}_{AD}$  & $G^{\mu\nu A}_B \overline{d}_C \gamma_{\mu}D_{\nu} \psi^D, \quad G^{\mu\nu A}_B D_{\nu} \overline{d}_C \gamma_{\mu} \psi^D$ & ($3,1,-1/3$) & $\Cap$ \tabularnewline 
\hline
${C_{8,8}}^{BD}_{AC}$ & $G^{\mu\nu A}_B \overline{L}^i \gamma_{\mu}D_{\nu} \psi^C_{iD}, \quad G^{\mu\nu A}_B D_{\nu} \overline{L}^i \gamma_{\mu} \psi^C_{iD}$ & ($8,2,-1/2$) & $\diamond$ \tabularnewline 
\hline
${C_{8,8}}^{BD}_{AC}$ & $G^{\mu\nu A}_B \overline{\ell} \gamma_{\mu}D_{\nu} \psi^C_D, \quad G^{\mu\nu A}_B D_{\nu} \overline{\ell} \gamma_{\mu} \psi^C_D$ & ($8,1,-1$) & $\diamond$ \tabularnewline 
\hline
${C_{\overline{3},3}}^{A}_B$ & $W^{\mu\nu i}_j D_\nu \overline{Q_L}_A^k \gamma_{\mu} \psi_{ik}^{jB}$ & ($3,4,1/6$) & $\star$ \tabularnewline 
\hline
${C_{\overline{3},3}}^{A}_B$  & $W^{\mu\nu i}_j D_\nu \overline{Q_L}^j_A \gamma_{\mu} \psi^B_i$ & ($3,2,1/6$) & $\star$ \tabularnewline 
\hline
${C_{\overline{3},3}}^{A}_B$  & $B^{\mu\nu} D_\nu \overline{Q_L}^i_A \gamma_{\mu} \psi_i^B$ & ($3,2,1/6$) & $\star$ \tabularnewline 
\hline
${C_{\overline{3},3}}^{A}_B$  & $W^{\mu\nu i}_j D_\nu \overline{u}_A \gamma_{\mu} \psi^{jB}_i$ & ($3,3,2/3$) & $\star$ \tabularnewline 
\hline
${C_{\overline{3},3}}^{A}_B$  & $B^{\mu\nu} D_\nu \overline{u}_A \gamma_{\mu} \psi^B$ & ($3,1,2/3$) & $\star$ \tabularnewline 
\hline
${C_{\overline{3},3}}^{A}_B$  & $W^{\mu\nu i}_j D_\nu \overline{d}_A \gamma_{\mu} \psi^{jB}_i$ & ($3,3,-1/3$) & $\star$ \tabularnewline 
\hline
${C_{\overline{3},3}}^{A}_B$  & $B^{\mu\nu} D_\nu \overline{d}_A \gamma_{\mu} \psi^B$ & ($3,1,-1/3$) & $\star$ \tabularnewline 
\hline
\end{tabular}
\end{center}
\caption{\label{tab:13}Dimension 6 operators of type $F D\psi f$.}
\end{table}

Within Table~\ref{tab:13}, there are a few different LEX SU(3) representations. Operators with an SU(3)$_\text{c}$ field strength tensor and a quark result in a LEX state with color charge dictated by Eq.~\ref{eq:3}. Meanwhile, in operators with only a gluon field strength tensor, the LEX state will be a color octet. In operators with an anti-quark as the only color-charged SM state, the LEX state will be a color triplet. For operators without the electroweak field strength tensor, the BSM state will be either an SU(2)$_\text{L}$ doublet or singlet, in the case of left- or right-handed SM fermions, respectively. LEX states in operators with the electroweak field strength tensor and a right-handed SM quark will be SU(2)$_\text{L}$ triplets. Finally, operators containing both the electroweak field strength tensor and a left-handed SM quark also contain LEX states that are electroweak doublets or quadruplets.

\subsubsection{Higher spin LEX representations in $FDf\psi$}

To compute the allowed higher spin LEX representations we need need a the Lorentz tensor product
\begin{align}
\begin{split}
[(1,0)\oplus(0,1)]&\otimes(1/2,1/2)\otimes[(1/2,0)\oplus(0,1/2)] \\
 &= [(2,1/2)\oplus(1/2,2)]\oplus 2[(1,1/2)\oplus(1/2,1)] \oplus [(1,3/2)\oplus(3/2,1)] \\ &~~~~\oplus [(3/2,0)\oplus(0,3/2)]\oplus2[(1/2,0)\oplus(0,1/2)]
 \end{split}
\end{align}

\begin{itemize}
\item $[(3/2,0)\oplus(0,3/2)]$ LEX representation has operators with Lorentz structure
\begin{align}
F_{\mu\nu}D^{\sigma}\overline{f}\gamma_{\sigma}\Psi^{[\mu\nu]},~~
F_{\mu\nu}D^{\nu}\overline{f}\gamma_{\sigma}\Psi^{[\sigma\mu]}, ~~~F_{\mu\nu}D_{\sigma}\overline{f}\gamma_{\mu}\Psi^{[\sigma\nu]}
\end{align}

\item $[(1,1/2)\oplus(1/2,1)]$ LEX states have operators with Lorentz structure  
\begin{align}
F_{\mu\nu}D^{\nu}\overline{f}\Psi^{\mu},~~ F_{\mu\nu}D^{\rho}\overline{f}\sigma^{\mu\nu}\Psi^{\rho}
\end{align}

\item $[(3/2,1)\oplus(1,3/2)]$ LEX representation has operators with Lorentz structure 
\begin{align}
F_{\mu\nu}D^{\nu}\overline{f}\gamma_{\sigma}\Psi^{\mu\sigma}
\end{align}

\item $[(2,1/2)\oplus(1/2,2)]$ LEX representation has operators with Lorentz structure 
\begin{align}
F_{\mu\nu}D_{\rho}\overline{f}\gamma_{\sigma}\Psi^{\mu\nu\rho\sigma}
\end{align}
\end{itemize}

\subsection{Operators of form $DD\psi f$ and $DD\psi f H$}
The next type of operator that we will consider includes a fermionic LEX state and a SM fermion, in addition to two covariant derivatives. We will consider dimension five operators with only this list of operator content, and dimension six operators which also include a Higgs field. The two gluons required for the $g$-$g$ portal are generated by the derivatives. As such, the SM fermion in operators of these types must be a quark, in order to introduce non-singlet SU(3)$_\text{c}$ charge. It is clear that this requirement also holds in the gluon-quark portal.

Following the same logic as in a previous section, the operators here that include the $D_\mu D_\nu \sigma^{\mu\nu}$ derivative structure are redundant with those of type $F \psi f $ or $F \psi f H$. Furthermore, the $D_\mu D_\nu \gamma^\mu \gamma^\nu$ operator form is once again redundant with those of type $D_\mu D^\mu$ and type $D_\mu D_\nu \sigma^{\mu\nu}$. As such, we may now choose to discard either operators with the derivative structure $D_\mu D_\nu \gamma^\mu \gamma^\nu$ or operators with the structure $D_\mu D^\mu$.

We start by examining operators of this type from within the $g$-$g$ portal. Within this portal, both derivatives are required to act on a color-charged state in order to generate two gluons. For simplicity, we choose to work with the $D_\mu D^\mu$ derivative structure. In operators with no Higgs fields, we have three possible operators, related by two integration by parts identities. As such, we can choose to only keep operators where both derivatives act on the SM fermion. While one may at first assume that there will be additional operators in the case that the operator contains a Higgs field, we note that no derivatives in operators of type $DD\psi f H$ may act on the Higgs within the $g$-$g$ portal. As such, there will still only be one independent operator of this type. Operators of this type are shown in Table~\ref{tab:14} for spin 1/2 particles. 
Within these operators, the LEX state is always a color anti-triplet. The LEX state can be a, SU(2) singlet, doublet, or triplet, in the same way as was explained for operators of type $fD\psi $ and $fD\psi H$.
\begin{table}[h!]
\begin{center}
\begin{tabular}{|c|c|c|c|}
\hline 
Dimension & Operators & (SU(3)$_{\text{c}}$, SU(2)$_{\text{L}}$, U(1)$_Y$) & Portal\tabularnewline
\hline 
\hline 
dim 5 & $D^2 \overline{Q_L}\psi$ & ($3,2,1/6$) & $\diamond$ \tabularnewline 
\hline  
dim 5 & $D^2 \overline{u}\psi$ & ($3,1,2/3$)  & $\diamond$ \tabularnewline 
\hline 
dim 5 & $D^2 \overline{d}\psi$ & ($3,1,-1/3$)  & $\diamond$ \tabularnewline 
\hline 
dim 6 & $D^2 \overline{Q_L}\psi H$ & ($3,3,-1/3$) & $\diamond$ \tabularnewline 
\hline
dim 6 & $D^2 \overline{Q_L}\psi H$ & ($3,1,-1/3$) & $\diamond$ \tabularnewline 
\hline
dim 6 & $D^2 \overline{u}\psi H$ & ($3,2,1/6$) & $\diamond$ \tabularnewline 
\hline
dim 6 & $D^2 \overline{d}\psi H$ & ($3,2,-5/6$) & $\diamond$ \tabularnewline 
\hline
dim 6 & $D^2 \overline{Q_L}\psi H^\dagger$ & ($3,3,2/3$) & $\diamond$ \tabularnewline 
\hline
dim 6 & $D^2 \overline{Q_L}\psi H^\dagger$ & ($3,1,2/3$) & $\diamond$ \tabularnewline 
\hline
dim 6 & $D^2 \overline{u}\psi H^\dagger$ & ($3,2,7/6$) & $\diamond$ \tabularnewline 
\hline
dim 6 & $D^2 \overline{d}\psi H^\dagger$ & ($3,2,1/6$) & $\diamond$ \tabularnewline 
\hline
\end{tabular}
\end{center}
\caption{\label{tab:14}Operators of type $DD\psi f$ and $DD\psi f H$ for spin 1/2 LEX states. SU(3)$_\text{c}$ indices are not written out. Operator dimension is included.}
\end{table}

Next, we consider operators of this type within the $g$-$q$ portal. In this case, we choose to work with the $D_\mu D_\nu \gamma^\mu \gamma^\nu$ derivative structure. Within this portal, for operators of type $DD\psi f$, there are again three operators and two integration by parts relations. Choosing the operator of type $\slashed{D} \psi \slashed{D} f$, we can implement fermionic equations of motion; operators of this type are redundant with those of type $Df\psi H$, discussed in Section~\ref{s3.2}. We also consider the operators including a Higgs field, of type $DD\psi f H$. In this case, there would be five operators in the portal, related by three integration by parts relations; this leaves two independent operators. We can choose these to be the operators $\slashed{D} \slashed{D} f \psi H$ and $\slashed{D} f \slashed{D} \psi H$. A fermionic equation of motion can also be applied in this case, yielding operators of type $Df\psi HH$, as also discussed in Section~\ref{s3.2}. As such, there are no operators of this type in the gluon-quark portal.

\subsection{Operators of form $DDD\psi f$}
The last operator type in the gluon-gluon portal up to dimension six contains three covariant derivatives, a SM fermion, and a fermionic LEX state. The SM fermion must be a quark in order to introduce a non-singlet SU(3)$_\text{c}$ charge. However, this operator type has already been fully discussed in our previous work  on the $W$-quark portal . The exact arguments from \cite{Carpenter:2025cys} also apply here; all operators of this type are redundant with each other and operators of types $FQ\psi$ and $Q\psi F H$ (from Section 3.1), $QD\psi F$ (from Section 3.2), and $DDQ\psi H$ and $DDQ\psi$ (from Section 3.3). Each of these operator types can be found in both the $g$-$g$ and $g$-$q$ portals; as such, we have no new operators of this type.

\subsection{Some Dimension 7 Operators: Operators of form $GF\psi f$}
The dimension seven operator type which will allow the largest possible SM quantum numbers is $GF\psi f$. As such, we enumerate operators of this type in Table~\ref{tab:15} and Table~\ref{tab:16}.
\begin{table}[h!]
\begin{center}
\begin{tabular}{|c|c|c|c|}
\hline 
Dimension & Operators & (SU(3)$_{\text{c}}$, SU(2)$_{\text{L}}$, U(1)$_Y$) & Portal \tabularnewline
\hline 
\hline 
dim 7 & $G_{\mu\nu} G^{\mu\nu} \overline{Q_L} \psi $ & ($42,2,1/6$) & $\Cap$ \tabularnewline 
\hline  
dim 7 & $G_{\mu\nu} G^{\mu\nu} \overline{Q_L} \psi $ & ($\overline{24},2,1/6$) & $\Cap$ \tabularnewline 
\hline  
dim 7 & $G_{\mu\nu} G^{\mu\nu} \overline{Q_L} \psi $ & ($15,2,1/6$) & $\Cap$ \tabularnewline 
\hline  
dim 7 & $G_{\mu\nu} G^{\mu\nu} \overline{Q_L} \psi $ & ($\overline{6},2,1/6$) & $\Cap$ \tabularnewline 
\hline  
dim 7 & $G_{\mu\nu} G^{\mu\nu} \overline{Q_L} \psi $ & ($3,2,1/6$) & $\Cap$ \tabularnewline 
\hline  
dim 7 & $G_{\mu\nu} G^{\mu\nu} \overline{u} \psi $ & ($42,1,2/3$) & $\Cap$ \tabularnewline 
\hline  
dim 7 & $G_{\mu\nu} G^{\mu\nu} \overline{u} \psi $ & ($\overline{24},1,2/3$) & $\Cap$ \tabularnewline 
\hline  
dim 7 & $G_{\mu\nu} G^{\mu\nu} \overline{u} \psi $ & ($15,1,2/3$) & $\Cap$ \tabularnewline 
\hline  
dim 7 & $G_{\mu\nu} G^{\mu\nu} \overline{u} \psi $ & ($\overline{6},1,2/3$) & $\Cap$ \tabularnewline 
\hline  
dim 7 & $G_{\mu\nu} G^{\mu\nu} \overline{u} \psi $ & ($3,1,2/3$) & $\Cap$ \tabularnewline 
\hline  
dim 7 & $G_{\mu\nu} G^{\mu\nu} \overline{d} \psi $ & ($42,1,-1/3$) & $\Cap$ \tabularnewline 
\hline  
dim 7 & $G_{\mu\nu} G^{\mu\nu} \overline{d} \psi $ & ($\overline{24},1,-1/3$) & $\Cap$ \tabularnewline 
\hline  
dim 7 & $G_{\mu\nu} G^{\mu\nu} \overline{d} \psi $ & ($15,1,-1/3$) & $\Cap$ \tabularnewline 
\hline  
dim 7 & $G_{\mu\nu} G^{\mu\nu} \overline{d} \psi $ & ($\overline{6},1,-1/3$) & $\Cap$ \tabularnewline 
\hline  
dim 7 & $G_{\mu\nu} G^{\mu\nu} \overline{d} \psi $ & ($3,1,-1/3$) & $\Cap$ \tabularnewline 
\hline  
\end{tabular}
\end{center}
\caption{\label{tab:15}Dimension 7 operators of type $GG\psi Q$ for spin 1/2 LEX fermions.}
\end{table}

\begin{table}[h!]
\begin{center}
\begin{tabular}{|c|c|c|c|}
\hline 
Dimension & Operators & (SU(3)$_{\text{c}}$, SU(2)$_{\text{L}}$, U(1)$_Y$) & Portal \tabularnewline
\hline 
\hline 
dim 7 & $G_{\mu\nu} G^{\mu\nu} \overline{L} \psi $ & ($27,2,-1/2$) & $\diamond$ \tabularnewline 
\hline  
dim 7 & $G_{\mu\nu} G^{\mu\nu} \overline{L} \psi $ & ($\overline{10},2,-1/2$) & $\diamond$ \tabularnewline 
\hline  
dim 7 & $G_{\mu\nu} G^{\mu\nu} \overline{L} \psi $ & ($10,2,-1/2$) & $\diamond$ \tabularnewline 
\hline  
dim 7 & $G_{\mu\nu} G^{\mu\nu} \overline{L} \psi $ & ($8,2,-1/2$) & $\diamond$ \tabularnewline 
\hline  
dim 7 & $G_{\mu\nu} G^{\mu\nu} \overline{L} \psi $ & ($1,2,-1/2$) & $\diamond$ \tabularnewline 
\hline  
dim 7 & $G_{\mu\nu} G^{\mu\nu} \overline{\ell} \psi $ & ($27,1,-1$) & $\diamond$ \tabularnewline 
\hline  
dim 7 & $G_{\mu\nu} G^{\mu\nu} \overline{\ell} \psi $ & ($\overline{10},1,-1$) & $\diamond$ \tabularnewline 
\hline  
dim 7 & $G_{\mu\nu} G^{\mu\nu} \overline{\ell} \psi $ & ($10,1,-1$) & $\diamond$ \tabularnewline 
\hline  
dim 7 & $G_{\mu\nu} G^{\mu\nu} \overline{\ell} \psi $ & ($8,1,-1$) & $\diamond$ \tabularnewline 
\hline  
dim 7 & $G_{\mu\nu} G^{\mu\nu} \overline{\ell} \psi $ & ($1,1,-1$) & $\diamond$ \tabularnewline 
\hline  
dim 7 & $G_{\mu\nu} W^{\mu\nu} \overline{Q_L} \psi $ & ($15,4,1/6$) & $\Cap$ \tabularnewline 
\hline  
dim 7 & $G_{\mu\nu} W^{\mu\nu} \overline{Q_L} \psi $ & ($\overline{6},4,1/6$) & $\Cap$ \tabularnewline 
\hline  
dim 7 & $G_{\mu\nu} W^{\mu\nu} \overline{Q_L} \psi $ & ($3,4,1/6$) & $\Cap$ \tabularnewline 
\hline  
dim 7 & $G_{\mu\nu} W^{\mu\nu} \overline{Q_L} \psi $ & ($15,2,1/6$) & $\Cap$ \tabularnewline 
\hline  
dim 7 & $G_{\mu\nu} W^{\mu\nu} \overline{Q_L} \psi $ & ($\overline{6},2,1/6$) & $\Cap$ \tabularnewline 
\hline  
dim 7 & $G_{\mu\nu} W^{\mu\nu} \overline{Q_L} \psi $ & ($3,2,1/6$) & $\Cap$ \tabularnewline 
\hline  
dim 7 & $G_{\mu\nu} W^{\mu\nu} \overline{u} \psi $ & ($15,3,2/3$) & $\Cap$ \tabularnewline 
\hline
dim 7 & $G_{\mu\nu} W^{\mu\nu} \overline{u} \psi $ & ($\overline{6},3,2/3$) & $\Cap$ \tabularnewline 
\hline
dim 7 & $G_{\mu\nu} W^{\mu\nu} \overline{u} \psi $ & ($3,3,2/3$) & $\Cap$ \tabularnewline 
\hline
dim 7 & $G_{\mu\nu} W^{\mu\nu} \overline{d} \psi $ & ($15,3,-1/3$) & $\Cap$ \tabularnewline 
\hline  
dim 7 & $G_{\mu\nu} W^{\mu\nu} \overline{d} \psi $ & ($\overline{6},3,-1/3$) & $\Cap$ \tabularnewline 
\hline  
dim 7 & $G_{\mu\nu} W^{\mu\nu} \overline{d} \psi $ & ($3,3,-1/3$) & $\Cap$ \tabularnewline 
\hline  
dim 7 & $G_{\mu\nu} W^{\mu\nu} \overline{L} \psi $ & ($8,4,-1/2$) & $\diamond$ \tabularnewline 
\hline  
dim 7 & $G_{\mu\nu} W^{\mu\nu} \overline{L} \psi $ & ($8,2,-1/2$) & $\diamond$ \tabularnewline 
\hline  
dim 7 & $G_{\mu\nu} W^{\mu\nu} \overline{\ell} \psi $ & ($8,3,-1$) & $\diamond$ \tabularnewline 
\hline  
dim 7 & $G_{\mu\nu} B^{\mu\nu} \overline{Q_L} \psi $ & ($15,2,1/6$) & $\Cap$ \tabularnewline 
\hline  
dim 7 & $G_{\mu\nu} B^{\mu\nu} \overline{Q_L} \psi $ & ($\overline{6},2,1/6$) & $\Cap$ \tabularnewline 
\hline  
dim 7 & $G_{\mu\nu} B^{\mu\nu} \overline{Q_L} \psi $ & ($3,2,1/6$) & $\Cap$ \tabularnewline 
\hline  
dim 7 & $G_{\mu\nu} B^{\mu\nu} \overline{u} \psi $ & ($15,1,2/3$) & $\Cap$ \tabularnewline 
\hline 
dim 7 & $G_{\mu\nu} B^{\mu\nu} \overline{u} \psi $ & ($\overline{6},1,2/3$) & $\Cap$ \tabularnewline 
\hline 
dim 7 & $G_{\mu\nu} B^{\mu\nu} \overline{u} \psi $ & ($3,1,2/3$) & $\Cap$ \tabularnewline 
\hline 
dim 7 & $G_{\mu\nu} B^{\mu\nu} \overline{d} \psi $ & ($15,1,-1/3$) & $\Cap$ \tabularnewline 
\hline 
dim 7 & $G_{\mu\nu} B^{\mu\nu} \overline{d} \psi $ & ($\overline{6},1,-1/3$) & $\Cap$ \tabularnewline 
\hline 
dim 7 & $G_{\mu\nu} B^{\mu\nu} \overline{d} \psi $ & ($3,1,-1/3$) & $\Cap$ \tabularnewline 
\hline 
dim 7 & $G_{\mu\nu} B^{\mu\nu} \overline{L} \psi $ & ($8,2,-1/2$) & $\diamond$ \tabularnewline 
\hline  
dim 7 & $G_{\mu\nu} B^{\mu\nu} \overline{\ell} \psi $ & ($8,1,-1$) & $\diamond$ \tabularnewline 
\hline  
\end{tabular}
\end{center}
\caption{\label{tab:16}Remaining dimension 7 operators of type $GF\psi f$ for spin 1/2 LEX fermions.}
\end{table}
 Operators of this type can be in either both the $g$-$q$ and $g$-$g$ portals or solely in the $g$-$g$ portal, based on whether the SM fermion is a quark or lepton.  We note for spin 1/2 LEX fermions  one can show that all possible Lorentz structures within this operator type are redundant with one another; we choose to use the Lorentz structure that is simplest in appearance, where the two field strength tensors  contract with one another.
 
Within operators of type $GF\psi f$, one may access LEX states that are singlets through quadruplets of SU(2). Operators without an SU(2) field strength tensor and no left-handed SM fermions yield electroweak singlet LEX states. Operators with an SU(2) field strength tensor and a right-handed SM fermion yield electroweak triplet LEX states. Finally, operators that include an SU(2) field strength tensor and a left-handed SM fermion contain electroweak doublet or quadruplet LEX states.

Finally, our discussion turns to the SU(3)$_\text{c}$ charges that one may reach in operators of type $GF\psi f$. The maximal case, where there are two SU(3)$_\text{c}$ field strength tensors and a SM antiquark, follows the algebra
\begin{align}
\begin{split}
    8\otimes8\otimes\overline{3} &= (\textcolor{red}{27}\oplus \textcolor{blue}{\overline{10}} \oplus \textcolor{orange}{10} \oplus \textcolor{Fuchsia}{8} \oplus \textcolor{Thistle}{8} \oplus \textcolor{Green}{1}) \otimes \overline{3} \\
    &= \textcolor{red}{\overline{42}\oplus 24 \oplus\overline{15}} \oplus \textcolor{blue}{\overline{15} \oplus \overline{15}} \oplus \textcolor{orange}{24\oplus 6} \oplus \textcolor{Fuchsia}{\overline{15}\oplus{6} \oplus \overline{3}} \oplus \textcolor{Thistle}{\overline{15}\oplus{6} \oplus \overline{3}} \oplus \textcolor{Green}{\overline{3}}.
\end{split}
\end{align}
From this, we see that we can reach 42-plets, 24-plets, 15-plets, sextets, and triplets of SU(3)$_\text{c}$. We note that the product $\overline{10}\otimes \overline{3}$ reaches two different kinds of $\overline{15}$, one with four lower SU(3)$_\text{c}$ indices and the other with one upper and two lower SU(3)$_\text{c}$ indices.  The 42-plet is an object with 2 upper indices and 3 lower indices: $\phi^{ij}_{klm}$.

The 42-plet could be produces with an associated quark or gluon at dimension seven from gluon fusion or quark-gluon fusion processes. 

Other dimension seven operator types with fermionic LEX states include  $G \psi f HH$, $F D\psi f H$, $DD\psi f HH$, $DDD\psi f$, $DDDD\psi f$. These can all potentially be found in both the $g$-$g$ and $g$-$q$ portals, depending on the specific SM field strength tensors or SM fermions that appear in the operators. Again, we do not comment in this work on whether any of these dimension seven operators types may be discarded due to redundancies.

\section{Some Phenomenological Examples}\label{s5}

As indication of the usefulness of this portal, we have chosen a few operators to investigate further. 

Though they are dimension seven operators, we have  chosen to delve into the three-gluon operators found in Table~\ref{tab:8}. Specifically, we have chosen to begin by investigate some of the simplest LEX operator the with $(1,1,0)$  singlet and the $(8,1,0)$ octet.  This $(1,1,0)$ operator is $f_{abc}G^{\mu\nu a} G^b_{\mu\sigma} G^{\sigma c}_{\nu } \phi$ (where the letters $a, b, c$ are SU(3)$_\text{c}$ adjoint indices).  In addition we have investigated  LEX color octet with the simplest color contraction operator. $(G^{\mu\nu } G_{\mu\sigma})_8 (G^{\sigma }_{\nu } \phi)_8$ . Recall from the previous section the three operators involving the scalar color octet have different color projections that are orthogonal to each other. In principle these have independent operator coefficients. This has allowed us to keep the simplest color contraction for the LEX octet.

We would like to note an issue with event simulation for colored exotics. One challenge we have run into in this work is that computational programs like MG5 and even Feynrules are not set up to simulate exotics in unusual SU(3)$_\text{c}$ representation. Implementing non-standard color contractions  has proven challenging. Using MG5 we were able to verify a few of the simpler color-spin selection rules derived in the previous section, by verifying the vanishing of  some interactions -but this was impeded by the group theory limitations on standard simulation tools.   

\begin{figure}[h!] 
    \centering
    \includegraphics[width=0.5\linewidth]{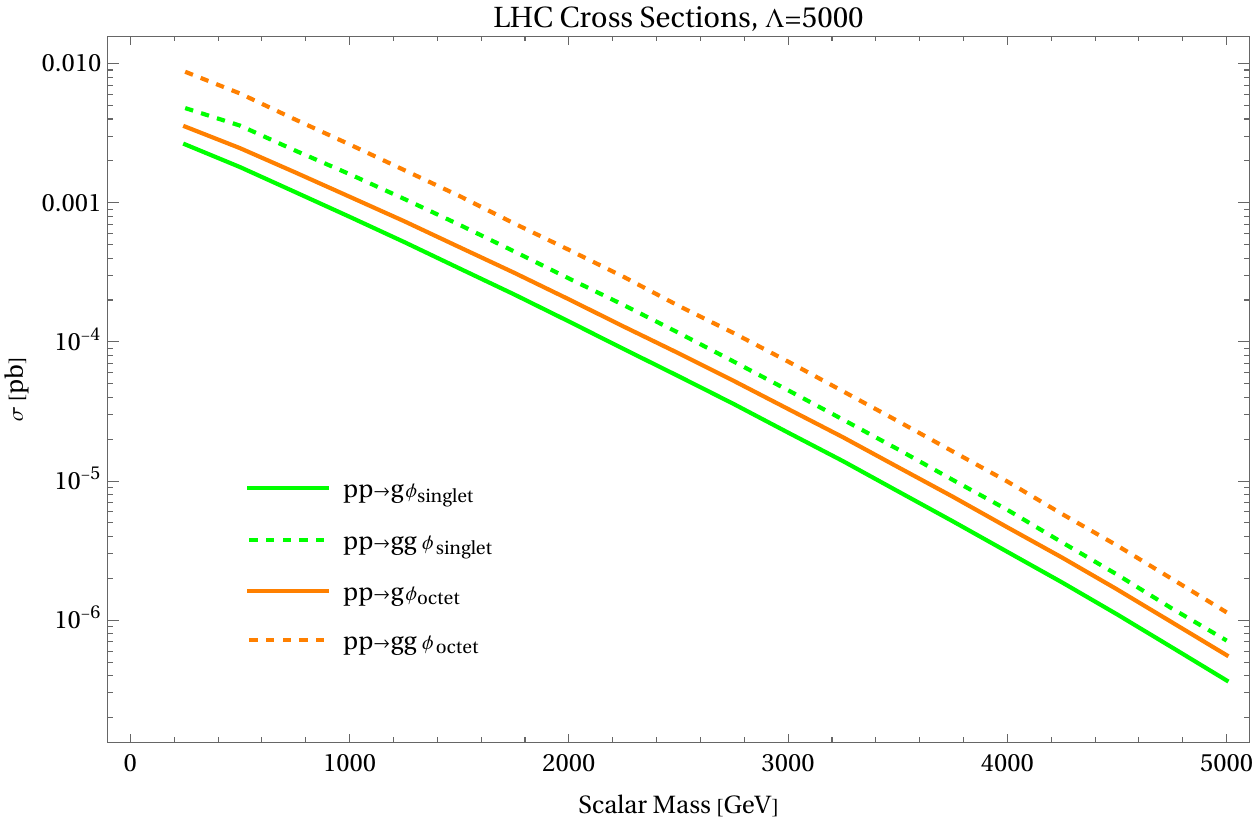}
    \caption{Cross sections from the color singlet and octet LEX state operators in Table~\ref{tab:8}.}
    \label{fig:6}
\end{figure}

We investigate the gluon fusion processes $gg\rightarrow g\phi$.  We utilized FeynRules to build a Universal Feynrules Output (UFO)\cite{Degrande:2011ua}, and fed this into MG5. We then used MG5 to generate 14 TeV LHC cross sections. These figures are shown in Fig.~\ref{fig:6}. We note that, at an effective cutoff value of 5 TeV, we have reasonably production cross section for 1 TeV LEX particles.

\begin{figure}[h!] 
    \centering
    \includegraphics[width=0.5\linewidth]{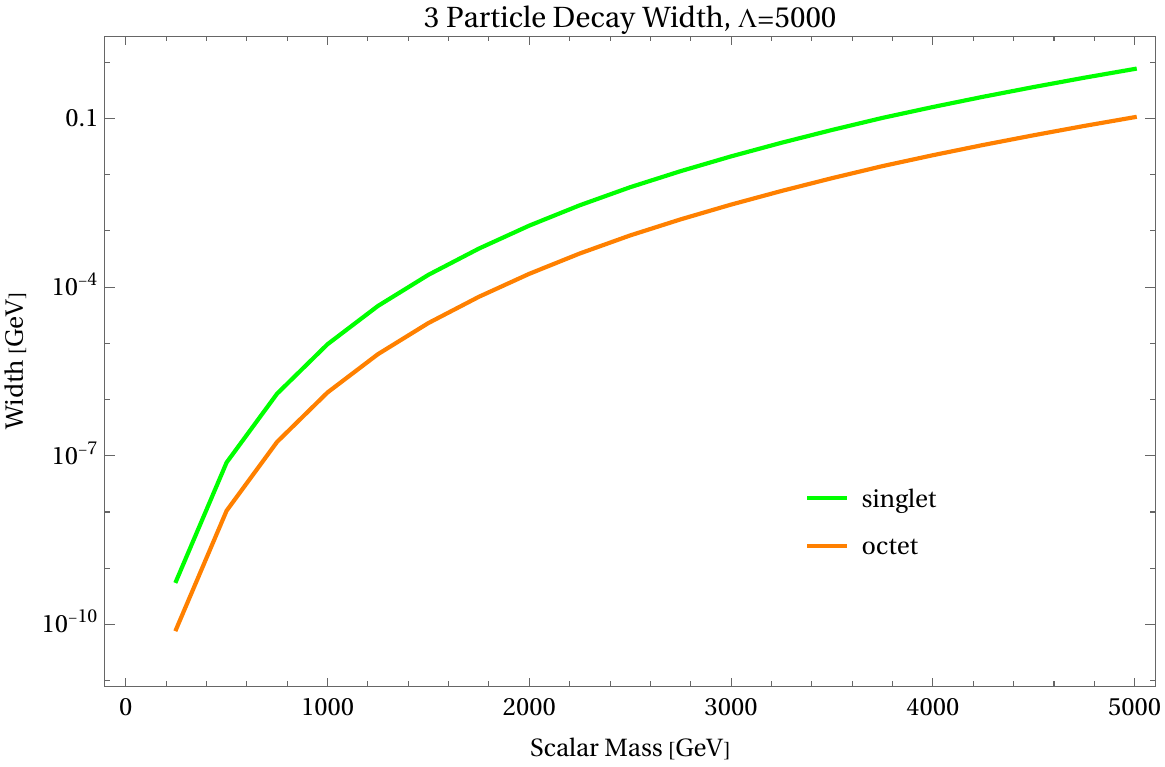}
    \caption{Three particle decay widths from the color singlet and octet LEX state operators in Table~\ref{tab:8}.}
    \label{fig:7}
\end{figure}

\begin{figure}[h!] 
    \centering
    \includegraphics[width=0.5\linewidth]{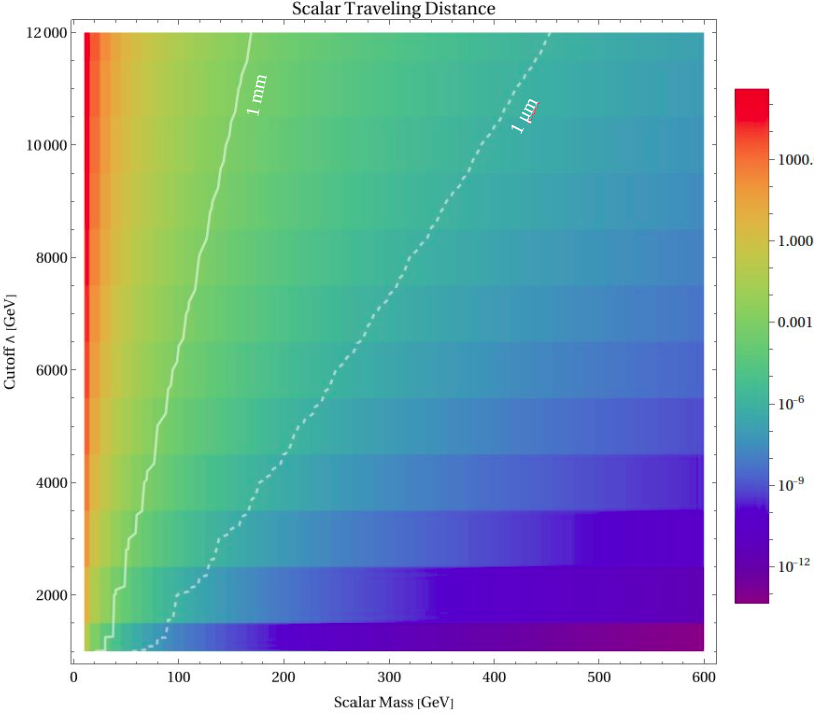}
    \caption{Traveling distance for the color singlet state.  }
    \label{fig:8}
\end{figure}

We also compute the decay widths of the LEX scalar, into three gluons $\Gamma(\phi\rightarrow g g g )$ shown in Fig.~\ref{fig:7}.
From this, we can find the decay length of our state, which are shown in Fig.~\ref{fig:8}.
In the figure we have made a contour plot of the decay length in the scalar mass vs effective cut-off plane.  This figure is made for LEX singlet $(1,1,0)$ states.  We do not expect harsh lower bounds on mass constraints of a neutral particle decaying to a three gluon resonance $\phi\rightarrow ggg$. Therefore, we find it reasonable to study LEX singlets in the 100's of GeV mass range. Here we find particle decay lengths between a millimeter and  a micro-meter for LEX states in the 100--200  GeV mass range and effective operator cut-off of 5 TeV or so.  If the singlet state can be produced through an alternative process This opens the interesting possibility of long-lived LEX states decaying through interesting dimension seven operators. Production and lifetime investigations may be interesting for the other dimension seven LEX operators.

\section{Conclusion}\label{s6}
In this work we have detailed all possible exotic states that singly couple to at least 2 gluons or at least on quark and one gluon at dimension six. Through this phenomenological portal we may specify all the exotic states that can be singly produced in specific production processes in a hadron collider processes-most notably in gluon fusion and quark-gluons fusion processes. We see that in these processes we can access particles of extremely exotic standard model charges, especially states in higher dimensional color representations. We see that---including some dimension seven operators---we can access light exotic states in 24, 27, 42, 35, and 64 dimensional color representations. 
Our operators also guarantee decays for light exotic states, which we have also discussed. 

We have mapped out some of the possible event topologies for single production of exotic states. We have also discussed some other phenomenological features for LEX operators such as color and spin selection rules.  Finally we have computed some LHC production and decay rates for a few of our more interesting operators.
It is our hope that this mapping of exotic terrain opens up the field to consideration of more exotic types of particle. 

\subsection*{Acknowledgments}
This work was partially supported by the OSU department of Physics.

We would also like to thank Dr. Guido Festuccia for several discussions on group theory.

\subsection*{Statement of Human Authenticity}

The authors affirm that all text, images, and
conceptual creation put forward in this work are
entirely human generated, with no input from generative
artificial intelligence (AI). The authors do not consent
to the use of this work in the training of generative AI
models.

\bibliographystyle{JHEP}

\bibliography{biblq.bib}

\end{document}